\documentclass[3p,12pt]{elsarticle}

\usepackage{lineno}
\modulolinenumbers[5]

\usepackage{amsfonts,amsmath,amssymb,graphicx}
\usepackage{color,mathrsfs,bm,appendix,subfigure}
\usepackage[dvipsnames]{xcolor}

\usepackage[figuresright]{rotating}
\usepackage{multirow,soul}
\usepackage{colortbl}

\usepackage{epstopdf}
\usepackage{capt-of}

\usepackage{booktabs}

\journal{Journal of Computational Physics}

\begin{document}

\begin{frontmatter}

\title{Implicit Discontinuous Galerkin Method for the Boltzmann Equation}

\author{Wei Su, Peng Wang, Yonghao Zhang, Lei Wu\corref{mycorrespondingauthor}}
\address{James Weir Fluids Laboratory, Department of Mechanical and Aerospace Engineering, \\University of Strathclyde, G1 1XJ Glasgow, United Kingdom}
\cortext[mycorrespondingauthor]{Corresponding author: lei.wu.100@strath.ac.uk}

\begin{abstract}

An implicit high-order discontinuous Galerkin (DG) method is developed to find steady-state solution of rarefied gas flow described by the Boltzmann equation with full collision operator. In the physical space, velocity distribution function is approximated by the piecewise polynomials of degree up to 4, while in the velocity space the fast spectral method is incorporated into the DG discretization to evaluate the collision operator. A specific polynomial approximation for the collision operator is proposed to reduce the computational complexity of the fast spectral method by $K$ times, where for two-dimensional problems $K$ is 15 when DG with 4th-order polynomials are used on triangular mesh. Based on the first-order upwind scheme, a sweeping technique is employed to solve the local linear equations resulting from the DG discretization sequentially over spatial elements. This technique can preserve stability of the scheme and requires no nonlinear limiter in solving hypersonic rarefied gas flow when the flow is fully resolved. Moreover, without assembling large sparse linear system, the computational cost in terms of memory and CPU time can be significantly reduced. Five different one/two-dimensional tests including low-speed microscale flows  and hypersonic rarefied gas flows are used to verify the proposed approach. Our results show that, DG schemes of different order of approximating polynomial require the same number of iterative steps to obtain the steady-state solution with the same order of accuracy; and the higher order the scheme, fewer spatial elements thus less CPU time is needed. Besides, our method can be faster than the finite difference solver by about one order of magnitude. The produced solutions can be used as benchmark data for assessing the accuracy of other gas kinetic solvers for the Boltzmann equation and gas kinetic models that simplify the Boltzmann collision operator.

\end{abstract}

\begin{keyword}
discontinuous Galerkin, high-order discretization, Boltzmann equation, fast spectral method, implicit scheme
\end{keyword}

\end{frontmatter}


\section{Introduction}

In gas kinetic theory, the motion of molecules in dilute gas is mathematically described by the one-particle velocity distribution function and macroscopic flow properties are derived from its velocity moments~\cite{Chapman1970}. Nowadays, this theory has been used for the description of transport phenomena in a wide range of scientific disciplines and applications such as the aerothermal dynamics in aerospace engineering, fusion processes in nuclear science, natural gas recovery and extraction in unconventional gas industry, freeze drying techniques in pharmaceutical and food manufactures, electron transport in semiconductor devices, and physics of diffuse matter in interstellar medium, just to name a few. In the Boltzmann's description, the variation of velocity distribution function comes from the linear streaming in the phase space, and the nonlinear interaction due to  binary collisions. Thus, the distribution function is a seven dimensional variable, with three in the physical space, three in the molecular velocity space, and one in the temporal space. Meanwhile, the nonlinear collision operator is a fivefold operator with three dimensions in the velocity space and two dimensions in a unit sphere (i.e. solid angle).

The multi-dimensional structure of the Boltzmann equation poses a real challenge to its numerical solution~\cite{Dimarco2014}. Historically, two major categories of approaches have been developed. One is the stochastic approach, which uses simulation particles to represent a large number of real molecules and mimic the molecular behaviors. The prevail one is the direct simulation Monte Carlo (DSMC) method developed by Bird~\cite{Bird1994}. During the simulation, particles move through the spatial space in a realistic manner with respect to the time, while intermolecular collisions and molecule-surface interactions are calculated in probabilistic manners. The other category is the deterministic approach, which adopts a numerical quadrature to approximate the integration with respect to the molecular velocity on a set of fixed discrete points~\cite{HUANG1967,Aristov2001}. As a result of discretization in the velocity space, the original kinetic equation is represented as a set of linear hyperbolic equations with nonlinear source terms that couple all the equations. To solve the resulting system, the usual schemes of traditional computational fluid dynamic (CFD) techniques for hyperbolic conservation laws can be applied straightforwardly for the streaming term. Some hybrid stochastic-deterministic approaches have also been proposed to solve the Boltzmann equation~\cite{Wagner1995,Buet1996,Baker2008}. Note that the majority of methods are based on the splitting technique, where the streaming and collision are treated separately.

In this paper, we focus on the deterministic method to solve the Boltzmann equation, which requires proper treatment of the linear streaming operator and nonlinear collision operator. The finite difference method (FDM), finite volume method  and finite element method have been successfully employed to approximate the derivatives with respect to the spatial ordinates~\cite{YANG1995323,KOLOBOV2007589,MIEUSSENS2000429,titarev2012,Gobbert2007,KITZLER20151539}. The advantage of these methods is that they have been well developed to achieve high order spatial and temporal accuracy. However, they might lose robustness and produce nonphysical solution, when the velocity distribution function has large variations and/or the kinetic equations become stiff. Another category of schemes is the semi-Lagrangian~\cite{CROUSEILLES20091730,CROUSEILLES20101927,GUCLU20123289} and Lagrangian methods~\cite{DIMARCO2013680,DIMARCO2013699}, which is designed to ensure positivity of the solutions. The basic idea of the schemes is to solve the streaming for the successive time steps by following the characteristics, i.e. molecular trajectories. The semi-Lagrangian methods still utilize fixed computational grid, but evaluate solution at the points that can be transported by the molecular velocity onto the computational grid within a time step. The Lagrangian methods update solution according to streaming without using a spatial mesh. Instead, the calculation reduces to a single manipulation for each discrete velocity.


For evaluating collision term, the most simple and widely used way is to replace the complicated collision operator by a relatively simple kinetic model, such as the Bhatnagar-Gross-Krook (BGK) model~\cite{BGK1954}, ellipsoidal statistical BGK model~\cite{ESBGK1966}, and Shakhov model~\cite{Shakhov1968}, which describes the relaxation of distribution function to the local equilibrium distributions obtained from macroscopic flow properties. Otherwise, the full Boltzmann collision operator should be calculated. The attempts to directly solve the full Boltzmann collision operator started from the late 1980s. Goldstein et al. constructed a discrete collision mechanics on the velocity nodes, which can preserve the main physical properties of the collision operator~\cite{Goldstein1989}. However, a large amount of discrete velocities are required, since post-collision velocities must fall on the grid points. The computational cost is of the order $O\left(\bar{N}^7\right)$ ($\bar{N}$ is the number of points in each velocity direction), and the nominal accuracy is less than first order in the velocity space~\cite{Bonylev1995}. Improvement by using an interpolation to map the post-collision velocities onto the velocity grid makes the performance of the scheme is comparable to or even faster than DSMC in normal shock wave simulation~\cite{Morris2008}. The kinetic theory group in Kyoto introduced another family of methods evaluating collision in the velocity space~\cite{Sone1989,Ohwada1993,Ohwada1996,KOSUGE200187}, in which the distribution function is expanded in terms of basis functions, while the collision operator is computed by the product of the expansion coefficients obtained at the discrete velocities and the numerical kernels that are the collision operators for the basis functions. The numerical kernels are pre-computed by numerical integration, which are restricted to the hard-sphere model and distribution function with cylindrical symmetry. Note that there are other schemes such as the projection method that evaluates the collision operator over a set of collision pairs with different velocities, aim distances and reflect angles~\cite{TCHEREMISSINE1998,Tcheremissine2006}, as well as method based on nodal-discontinuous Galerkin discretization of the collision operator and a bi-linear convolution of the Galerkin projection~\cite{Alekseenko2018}. 

Instead of directly calculating the collision integral on discrete velocities, there is another route to approximate collision in frequency domain using Fourier transform techniques. These methods can present accuracy of typical spectral approaches. Besides, they can reduce computational cost through fast spectral algorithm. The pioneering work was introduced by Bobylev for Maxwell molecules~\cite{Bobylev1988}. Then, several spectral methods were developed, which have computational cost of the order up to $O\left(\bar{N}^6\right)$~\cite{Pareschi1996,Bobylev1997}. Their computational cost can be reduce to $O\left(\bar{N}^3\log\bar{N}\right)$ for distribution function possessing cylindrical symmetry, when the fast Fourier transform (FFT) and Hankel transform are employed~\cite{Watachararuangwit2009}. However, the accuracy is only of $O\left(\bar{N}^{-1/2}\right)$. Based on the Carleman representation, an algorithm was developed for hard-sphere molecules to achieve accuracy of $O\left(\bar{N}^{-2}\right)$, where the integration over the unit sphere is separated from the one over the velocity space~\cite{BOBYLEV1999869}. By employing generalized Radon and X-ray transform, its computational cost is of $O\left(\bar{N}^6\log\bar{N}\right)$. The algorithm for variable hard-sphere molecules of accuracy $O\left(\bar{N}^{-2}\right)$ was also proposed with complexity of $O\left(\bar{N}^6\right)$~\cite{Ibragimov2002}. The fast spectral method (FSM) that is spectrally accurate has been developed since the new century~\cite{Pareschi2000,FILBET2003457}. By means of the Carleman representation, the method is improved with the computational cost reduced to $O\left(\bar{M}^2\bar{N}^3\log\bar{N}\right)$, where $\bar{M}$ is the number of polar and azimuthal angles~\cite{Filbet2006,Mouhot2006}. This is in general the fastest algorithm to data. To extend the applicability of the FSM, novel anisotropic collision kernels were designed and incorporated, which can deal with all inverse power-law potentials (except the Coulomb potential) as well as the Lennard-Jones potential~\cite{WU2013,wu2014}. Later, the collision kernel for Lennard-Jones potential was fully resolved, however, the computational cost increases to $O\left(\bar{M}^2\bar{N}^4\log\bar{N}\right)$~\cite{Wu2015}. This method has been successfully applied to solve many canonical rarefied gas flows, where the computational efficiency is much higher than the low-variance DSMC method for low-speed flows~\cite{wu2014,Radtke2011}. 

The challenge to numerically solve the Boltzmann equation with full collision operator is that the computational cost becomes immediately prohibitive for realistic problems, since: 1) the number of governing equations is large due to discretization in the velocity space; 2) for each equation, the collision operator needs to be evaluated at every spatial grid points or elements (even the Lagrangian methods need spatial mesh for approximation of collision). Therefore, high-order CFD approach is critical to improve efficiency of discretization in the spatial space, thus reduce the computational cost. One of the promising methods for this purpose is the discontinuous Galerkin (DG) method, which was first introduced for the neutron transport equation~\cite{Reed1973}. The DG method provides advantages including: achieving high-order of accuracy with relatively low effort, easy formulation for arbitrary geometry, straightforward implementation of boundary condition with the same high-order accuracy as in the interior of the computational domain, as well as the efficient implementation for parallelism and adaptive refinement. After combining an explicit high-order Runge-Kutta time marching scheme, the method has great success in solving convection-dominated problems~\cite{Cockburn1998,Cockburn2001}. The explicit DG method has been applied to solve the kinetic model equations~\cite{SU2015123}. Very recently, it has also been applied to the full Boltzmann equation with the variable soft-sphere collision kernel, in which the collision operator is calculated based on a FSM having a cost at the order of $O\left(\bar{M}^2\bar{N}^4\log\bar{N}\right)$~\cite{JAISWAL2018}. It has been shown that the second-order DG method is 15 times faster than the second-order finite volume scheme~\cite{SU2015123}. However, higher-order explicit DG scheme is not superior to the lower-order one, mainly due to the fact that the time step restricted by the Counrant-Friedrichs-Lewy (CFL) condition becomes extremely small~\cite{KUBATKO20089697}; thus the number of iteration becomes very large in finding steady-state solution for high-order discretization.

Note that the FSM has also been incorporated in the Boltzmann solver based on Lagrangian method for streaming~\cite{DIMARCO201846}. The solution from this method is currently limited to first-order accuracy in space and time. Besides, the time marching is an explicit scheme, thus the total number of time steps is still enormous to obtain a steady solution. It is also interesting to mention that there is a class of methods, named (discrete) unified gas-kinetic scheme, sharing some properties with the semi-Lagrangian scheme~\cite{LIU2016305,ZHU201616,guo2013discrete,guo2015discrete}, in which the flux transport across spatial cell interface contains the evolution of distribution function along the molecular trajectories within a time step due to both stream and collision processes. By coupling the evaluations of both streaming and collision, the scale of spatial discretization can be reduced. These methods are first developed based on kinetic model equations. Recently, approximation of the Boltzmann collision operator using the FSM is incorporated to correct the relaxation of distribution functions to the local equilibrium states beyond the continuum flow regime~\cite{LIU2016305}.  

In this paper, we represent a DG method to solve the full Boltzmann equation, which is devoted to improving the scheme in the following ways:

\begin{itemize}
	\item{Implicit iterative schemes are employed to relax the limitation on time step from  the CFL condition. As a result, the superiority of high-order discretization in the DG method can be demonstrated, which is in sharp contrast to the explicit DG where the CFL number is rather small.}

	\item{An novel scheme is proposed to reduce the computational complexity when using the FSM to calculate the collision operator, say, by 15 times when using 4th order approximating polynomials on two-dimensional triangular mesh.}

	\item{A strategy based on the sweeping technique is introduced, which can avoid solving large sparse linear system, and stabilize the scheme without using any nonlinear limiter when the rarefied gas flow is fully resolved.}
	
\end{itemize} 

The remainder of the paper is organized as follows. In Sec.~\ref{Boltzmann}, the Boltzmann equation and the FSM are introduced. In Sec.~\ref{DG}, the implicit DG method is described with details in the formulation of collision operator. A scheme to reduce the complexity of DG discretization for the collision operator is proposed in Sec.~\ref{ComputationComplex}, while the sweeping strategy to solve the linear systems is described in Sec.~\ref{Sweep}. In Sec.~\ref{Results}, five different problems including one-dimensional shock wave, two-dimensional hypersonic flow past a square cylinder, lid driven cavity flow and two thermal low-speed microscale flows are simulated to assess the accuracy and efficiency of the proposed scheme. Conclusions are presented in Sec.~\ref{Concludsion}.

\section{The Boltzmann Equation and the Fast Spectral Method}\label{Boltzmann}

In kinetic theory, the state of a gas system is described by the one-particle velocity distribution function $f\left(t,\bm x,\bm v\right)$, which is a function of the time $t$, the spatial position $\bm x=\left(x_1,x_2,x_3\right)$, and the molecular velocity $\bm v=\left(v_1,v_2,v_3\right)$. Neglecting the external force, the evolution of velocity distribution function for a single-species monatomic gas is governed by the following Boltzmann equation:
\begin{equation}
\frac{\partial f}{\partial t} + \bm v\cdot\frac{\partial f}{\partial\bm x}=\mathcal{C}\left(f,f_{*}\right),
\end{equation}
where $\mathcal{C}\left(f,f_{*}\right)$ is the Boltzmann collision operator that is usually split into the gain term $\mathcal{C}_{+}$ and loss term $\mathcal{C}_{-}$:
\begin{equation}\label{collision_term}
\mathcal{C}\left(f,f_{*}\right)=\underbrace{\int\int B\left(\theta,|\bm v-\bm v_{*}|\right)f\left(\bm v'_{*}\right)f\left(\bm v'\right)\mathrm{d}\Omega\mathrm{d}\bm v_{*}}_{\mathcal{C}_{+}}-\underbrace{\nu(\bm v) f(\bm v)}_{\mathcal{C}_{-}}.
\end{equation}
with the collision frequency
\begin{equation}\label{collision_frequency}
\nu(\bm v)=\int\int B\left(\theta,|\bm v-\bm v_{*}|\right)f\left(\bm v_{*}\right)\mathrm{d}\Omega\mathrm{d}\bm v_{*}.
\end{equation}

Note that here $B\left(\theta,|\bm v-\bm v_{*}|\right)$ is the collision kernel; $\bm v$, $\bm v_{*}$ are the pre-collision molecular velocities of a collision pair, and $\bm v'$, $\bm v'_{*}$ are the corresponding post-collision molecular velocities; $\Omega$ is the unit vector along the relative post-collision velocity $\bm v'-\bm v'_{*}$, while $\theta$ is the deflection angle between the pre- and post-collision relative velocities. For simplicity, the time and spatial position is omitted in writing the distribution function, collision operator, and collision frequency.

The velocity distribution function is defined such that $f\left(t,\bm x,\bm v\right)\mathrm{d}\bm x\mathrm{d}\bm v$ is the number of gas molecules in the phase-space volume $\mathrm{d}\bm x\mathrm{d}\bm v$. All macroscopic quantities, such as mass density $\rho$, bulk velocity $\bm u=\left(u_1,u_2,u_3\right)$, temperature $T$, pressure tension $\bm P$ and heat flux $\bm Q=\left(Q_1,Q_2,Q_3\right)$ can then be calculated via velocity moments of the distribution function. For simplicity, we use non-dimensional variables hereafter: $\bm x$ is normalized by a characteristic flow length $H$, $T$ is normalized by a reference temperature $T_0$, $\rho$ is normalized by the average density $\rho_0$ at $T_0$, $\bm v$ and $\bm u$ are normalized by the most probable speed $v_\text{m}=\sqrt{2k_\text{B}T_0/m}$ with $k_{\text{B}}$ and $m$ being the Boltzmann constant and molecular mass, respectively, $t$ is normalized by $H/v_\text{m}$, $f$ is normalized by $\rho_0/mv^3_\text{m}$, $\bm P$ is normalized by $\rho_0k_\text{B}T_0/m$, and $Q_i$ is normalized by $\rho_0k_\text{B}T_0v_\text{m}/m$. Therefore, we have 
\begin{equation}
\begin{aligned}[b]
\rho = \int f\mathrm{d}\bm v,\quad\bm u=\frac{1}{\rho}\int\bm vf\mathrm{d}\bm v,\quad T=\frac{2}{3\rho}\int|\bm v-\bm u|^2f\mathrm{d}\bm v,\\
\bm P=2\int\left(\bm v-\bm u\right)\otimes\left(\bm v-\bm u\right)f\mathrm{d}\bm v,\quad \bm Q=\int\left(\bm v-\bm u\right)|\bm v-\bm u|^2f\mathrm{d}\bm v.
\end{aligned}
\label{Macro}
\end{equation}


The collision kernel $B\left(\theta,|\bm v-\bm v_{*}|\right)$, depending on the modules of the pre-collision relative velocity and the deflection angle, is only determined when a certain intermolecular potential is given~\cite{Chapman1970}. The detailed structure of the collision kernel is usually very complicated, except that of the ideal hard-sphere molecule. In the history, both for the analytical and numerical convenience, specific simplification is adopted with the aim to recover the correct transport coefficients, which results in various molecular models that are widely used in the DSMC method. The key to these models is that transport coefficients such as shear viscosity, thermal conductivity, and diffuse coefficient are recovered over the temperature range considered. In this paper, the collision kernel is modeled as~\cite{Mouhot2006}:
\begin{equation}
B\left(\theta,|\bm v-\bm v_{*}|\right)=\frac{5|\bm v-\bm v_{*}|^{2\left(1-\omega\right)}}{2^{7-\omega}\Gamma\left(\frac{5-2\omega}{2}\right)Kn}\sin^{1-2\omega}\left(\frac{\theta}{2}\right),
\label{kernel}
\end{equation}
where $\Gamma$ is the Gamma function, $\omega$ is the viscosity index (i.e. the shear viscosity $\mu$ of the gas is proportional to $T^{\omega}$) and $Kn$ is the unconfined Knudsen number given at the reference condition:
\begin{equation}
Kn=\frac{\mu\left(T=T_0\right)}{\rho_0H}\sqrt{\frac{m\pi}{2k_{\text{B}}T_0}}.
\end{equation}

It is noted that the specific form~\eqref{kernel} introduced by Mouhot and Pareschi enables the development of Carleman-representation-based FSM to deterministically compute the collision operator. It has the ability to mimic the growth trend of collision kernel when decreasing the deflection angle and recover correct values of shear viscosity, however it cannot deal with general forms of soft potentials. By introduce another free-parameter into the collision kernel, the authors have extended the applicability of FSM to all inverse power law potentials (except the Coulomb potential), thus to recover the correct value of diffusion coefficient~\cite{WU2013,wu2014}. We also mention that more general collision models including the Lennard-Jones potential has been incorporated into the FSM~\cite{Wu2015,JAISWAL2018}. For general collision kernel, the computational cost will be one order of magnitude higher than that of Eq.~\eqref{kernel}; therefore, in this paper  Eq.~\eqref{kernel} is adopted to demonstrate efficiency and accuracy of the DG method on the spatial discretization. As a matter of fact, if viscosity index is chosen appropriately, the collision kernel can yield accurate results when compared to that of the realistic Lennard-Jones potential~\cite{wu2014,Wu2015}.

\subsection{The fast spectral method}

The Boltzmann collision operator~\eqref{collision_term} is a five-fold integral with three dimensions in the molecular velocity space and two dimensions in a unit sphere. In this paper, the FSM is applied to evaluate the collision operator, details of which can be found in~\cite{Mouhot2006,WU2013,Wu2015}.

Firstly, the distribution function is periodized on a truncated domain $\mathcal{D}=[-L,L]^3$ and expanded into Fourier series with $N_1\times N_2\times N_3$ components:
\begin{equation}
f\left(t,\bm x,\bm v\right)=\sum^{N/2-1}_{j=-N/2}\bar{f}^{j}\left(t,\bm x\right)\exp\left(\imath\bm\xi^{j}\cdot\bm v\right),
\end{equation}
\begin{equation}
\bar{f}^{j}\left(t,\bm x\right)=\frac{1}{\left(2L\right)^3}\int_{\mathcal{D}}f\left(t,\bm x,\bm v\right)\exp\left(-\imath\bm\xi^{j}\cdot\bm v\right)\mathrm{d}\bm v,
\label{Spectrum}
\end{equation}
where $L$ is the maximum truncated velocity, $\imath$ is the imaginary unit, $\bar{f}^{j}$ is the spectrum of the velocity distribution function, $\bm\xi^{j}=j\pi/L$ is the discrete frequency with $j=\left(j_1,j_2,j_3\right)$ and $N=\left(N_1,N_2,N_3\right)$ denoting the index and total number of frequencies. In order to take the advantage of FFT, the discretized frequency components are equally spaced.

Then, the gain term in collision operator and the collision frequency are evaluated through expanding in Fourier series:
\begin{equation}
\mathcal{C}_{+}=\sum_{j=-N/2}^{N/2-1}\bar{\mathcal{C}}^{j}_{+}\exp\left(\imath\bm\xi^{j}\cdot\bm v\right), \quad
\nu=\sum_{j=-N/2}^{N/2-1}\bar{\nu}^{j}\exp\left(\imath\bm\xi^{j}\cdot\bm v\right),
\label{spectrumColl0}
\end{equation}
where the $j$-th Fourier modes of the gain term in Eq.~\eqref{collision_term} and collision frequency~\eqref{collision_frequency} are calculated from the spectrum $\bar{f}$ as follows~\cite{WU2013,wu2014}:
\begin{equation}
\bar{\mathcal{C}}_{+}^{j}=\sum^{N/2-1}_{\substack{l+m=j\\l,m=-N/2}}\bar{f}^{l}\bar{f}^{m}\beta\left(l,m\right),\quad \bar{\nu}^{j}=\bar{f}^{j}\beta\left(j,j\right).
\label{spectrumColl}
\end{equation}
Here, $\beta$ is the collision kernel mode, whose $\left(l,m\right)$-th component is approximated through $M_\text{qua}$-point numerical quadrature in spherical coordinates as:
\begin{equation}\label{GL_kernel}
\begin{aligned}
\beta\left(l,m\right)\simeq\frac{20\sum^{M_\text{qua}}_{p,q=1}\sin\left(\theta_p\right)\Psi\left(\sqrt{|\bm\xi^{m}|^2-\left(\bm\xi^{m}\cdot\bm e_{p,q}\right)^2}\right)\Phi\left(\bm\xi^{l}\cdot\bm e_{p,q}\right)\varpi_p\varpi_q}{2^{7-\omega}\Gamma\left(\frac{5-2\omega}{2}\right)Kn},
\end{aligned}
\end{equation}
where $\theta_p$ ($\phi_q$) and $\varpi_p$ ($\varpi_q$) are the $p$ ($q$)-th point and weight of the quadrature rule, respectively, for $\theta$, $\phi\in\left[0,\pi\right]$, and $\bm e_{p,q}=\left(\sin\theta_p\cos\phi_q,\sin\theta_p\sin\phi_q,\cos\theta_p\right)$. The functions $\Psi$ and $\Phi$ are  $\Psi\left(a\right)=2\pi\int_0^R\rho^{1-\gamma}J_0\left(\rho a\right)\mathrm{d}\rho$ and $\Phi\left(a\right)=2\int_0^R\rho^{2\left(1-\omega\right)+\gamma}\cos\left(\rho a\right)\mathrm{d}\rho$,
where $J_0$ is the zeroth-order Bessel function, and $R$ is the radius of the sphere to support the distribution function, which is chosen approximately as $R=2\sqrt{2}L/(2+\sqrt{2})$ to avoid the aliasing error~\cite{WU2013}. Note that by estimating through numerical quadrature, frequencies $\bm\xi^m$ and $\bm\xi^l$ appear in two different functions in the final form of $\beta\left(l,m\right)$, thus Eq.~\eqref{spectrumColl} can be calculated by FFT-based convolution~\cite{WU2013}.



\section{Implicit Discontinuous Galerkin Method}\label{DG}


To obtain the stationary solution of the Boltzmann equation, the following implicit iterative scheme is usually applied:
\begin{equation}
\bar{\nu}f^{(t+1)} + \bm v\cdot\frac{\partial f^{(t+1)}}{\partial\bm x}=\bar{\nu}f^{(t)}+\mathcal{C}\left(f^{(t)},f^{(t)}_*\right),
\label{ITR1}
\end{equation}
where the superscripts $(t)$ and $(t+1)$ represent two consecutive iteration steps. The iteration is terminated when the convergence to the steady solution is achieved. The parameter $\bar{\nu}$ is a positive constant which is the reciprocal time step in the backward-Euler method and highly influences convergence property of the iterative scheme: too large (small) $\bar{\nu}$ results in slow convergence (numerical instability). Usually, to strike a balance between efficiency and stability of the iteration, $\bar{\nu}$ is chosen to be the order of mean collision frequency $\int\nu(\bm v)f(\bm v)d\bm v$. Therefore, a safe choice of $\bar{\nu}$ is the minimum mean collision frequency in the whole computational domain. However, one needs a good estimation for the minimum $\bar{\nu}$ before calculation.

Another way to find the steady-state solution is to neglect the derivative of distribution function with respect to the time, yielding $\bm v\cdot\partial f/\partial\bm x=\mathcal{C}$. Then, the collision frequency and gain term of the Boltzmann collision operator are evaluated based on the approximation of distribution at the iteration step $t$, while other terms are solved at the next iteration step by:
\begin{equation}
\nu^{(t)}(\bm{x},\bm{v})f^{(t+1)}(\bm{x},\bm{v}) + \bm v\cdot\frac{\partial f^{(t+1)}(\bm{x},\bm{v})}{\partial\bm x}=\mathcal{C}^{(t)}_{+}(\bm{x},\bm{v}),
\label{ITR2}
\end{equation}

In the following sections, we will denote the iterative scheme~\eqref{ITR1} with mean collision frequency as `ITR-MEAN' and the iterative scheme~\eqref{ITR2} with local collision frequency as `ITR-LOC'. The two iteration schemes can lead to different computational complexity and convergence history, which will be discussed in Sec.~\ref{Sweep}. For conciseness, we will omit the index of iteration step in the following unless necessary.

\subsection{DG formulation for the Boltzmann equation}

Now we present the DG method to find the steady-state solutions of rarefied gas flows described by~\eqref{ITR1} and~\eqref{ITR2}.
Let $\Delta\in\mathbb{R}^d$ be a computational domain in the $d$-dimensional spatial space with boundary $\partial\Delta$. Then, the domain is partitioned into $M_\text{el}$ disjoint regular elements $\Delta_i$. The DG method provides an approximate solution to the velocity distribution function $f$ on each element $\Delta_i$ in some piecewise finite element spaces $\mathcal{V}$ of the following form:
\begin{equation}
\mathcal{V}=\{\varphi_r\left(\bm x\right):\varphi_r|_{\Delta_i}\in\mathcal{P}^k\left(\Delta_i\right),\ r=1,\dots,K,\forall\Delta_i\in\Delta\},
\end{equation}
where $\mathcal{P}^k$ denotes the space of $k$-th order polynomials, thus we have
\begin{equation}
f\left(\bm x,\bm v\right)=\sum^{K}_{r=1}\varphi_r\left(\bm x\right)F_r\left(\bm v\right),
\label{polynomial}
\end{equation}
with $F_r$ being the degree of freedom for the distribution function. In general, the degrees of freedom are unknowns for which the equations are being solved. Together with the basis functions $\varphi_r$, they give the final polynomial estimation of $f$ within a spatial element $\Delta_i$. The number of degree of freedom, $K$, dependents on the shape of element employed. For example, $K=k+1$ for line segament in one-dimensional (1D) problem, and $K=\left(k+1\right)\left(k+2\right)/2$ for triangular element in two-dimensional (2D) problem.

In order to determine $F_r$, standard techniques of finite element formulations are applied to obtain the weak formulation of the governing system. Introducing $\left(\cdot\right)$ and $\langle\cdot\rangle$ as $\left(a,b\right)_{\Delta_i}=\int_{\Delta_i}\left(a\cdot b\right)\mathrm{d}\bm x$ and $\langle a,b\rangle_{\partial\Delta_i}=\int_{\partial\Delta_i}\left(a\cdot b\right)\mathrm{d}\Upsilon$ to denote operators on the element $\Delta_i$ and its boundary $\partial\Delta_i$, respectively, we find the approximation of distribution function satisfies the following equation (take the ITR-MEAN scheme~\eqref{ITR1} as an example):
\begin{equation}
-\left(\nabla\varphi_s,\bm vf\right)_{\Delta_i}+\langle\varphi_s,\hat{\bm H}\cdot\bm n\rangle_{\partial\Delta_i}+\left(\varphi_s,\bar{\nu} f\right)_{\Delta_i}=\left(\varphi_s,\mathcal{C}\right)_{\Delta_i}+\left(\varphi_s,\bar{\nu} f\right)_{\Delta_i},
\label{local}
\end{equation}
where $s=1,\dots,K$, $\bm n$ is the outward unit normal vector, and $\hat{\bm H}$ is the numerical flux that depends on the solutions from both sides of $\partial\Delta_i$, since the solution of $f$ is discontinuous there. We define the numerical flux from the first-order upwind principle as:
\begin{equation}
\hat{\bm H}\cdot\bm n=\frac{1}{2}\bm v\cdot\bm n\left(f+f_{\text{ext}}\right)+\frac{1}{2}|\bm v\cdot\bm n|\left(f-f_{\text{ext}}\right).
\label{flux}
\end{equation}
with $f_{\text{ext}}$ being the distribution from a neighboring element that shares the boundary $\partial\Delta_i$ with $\Delta_i$. If $\partial\Delta_i$ is at the boundary of computational domain, i.e. $\partial\Delta_i\cap\partial\Delta\neq0$, $f_{\text{ext}}$ is evaluated using the given boundary condition. 

Now, we focus on the formulation of $\left(\varphi_s,\mathcal{C}\right)_{\Delta_i}=\left(\varphi_s,\mathcal{C}_{+}\right)_{\Delta_i}-\left(\varphi_s,\nu f\right)_{\Delta_i}$ in Eq.~\eqref{local}. Inserting the polynomial expansion of distribution function~\eqref{polynomial} into Eq.~\eqref{Spectrum}, the $j$-th spectrum component of the distribution function can be rewritten in the following polynomial form:
\begin{equation}
\bar{f}^{j}\left(\bm x\right)=\sum^{K}_{r=1}\varphi_r\left(\bm x\right)\bar{F}^{j}_r,
\end{equation}
where $\bar{F}^{j}_r=\frac{1}{\left(2L\right)^3}\int_{\mathcal{D}}F_r\left(\bm v\right)\exp\left(-\imath\bm\xi^{j}\cdot\bm v\right)\mathrm{d}\bm v$ is the spectrum of the degree of freedom.

With some algebraic calculations from Eqs.~\eqref{spectrumColl0} and~\eqref{spectrumColl}, the DG discretization of the gain term of the Boltzmann collision operator and the collision frequency are expressed as
\begin{equation}
\mathcal{C}_{+}=\sum^{K}_{p=1}\sum^{K}_{r=1}\varphi_p\varphi_r\Xi_{p,r},\quad\nu=\sum^{K}_{p=1}\varphi_p\Lambda_p,
\end{equation}
where
\begin{equation}
\Xi_{p,r}=\sum_{j=-N/2}^{N/2-1}\sum^{N/2-1}_{\substack{l+m=j\\l,m=-N/2}}\bar{F}^{l}_p\bar{F}^{m}_r\beta\left(l,m\right)\exp\left(\imath\bm\xi^{j}\cdot\bm v\right),\quad
\Lambda_p=\sum_{j=-N/2}^{N/2-1}\bar{F}^{j}_p\beta\left(j,j\right)\exp\left(\imath\bm\xi^{j}\cdot\bm v\right).
\label{DoFColl}
\end{equation}

Finally, we obtain that
\begin{equation}
\left(\varphi_s,\mathcal{C}_{+}\right)_{\Delta_i}=\sum^{K}_{p=1}\sum^{K}_{r=1}\left(\varphi_s,\varphi_p\varphi_r\right)_{\Delta_i}\Xi_{p,r},
\label{Direct2}
\end{equation}
\begin{equation}
\left(\varphi_s,\nu f\right)_{\Delta_i}=\sum^{K}_{p=1}\sum^{K}_{r=1}\left(\varphi_s,\varphi_p\varphi_r\right)_{\Delta_i}\Lambda_pF_r,
\label{Direct1}
\end{equation}


\subsection{Discretization in the molecular velocity space}

In order to obtain the macroscopic flow properties~\eqref{Macro} and the spectrum~\eqref{Spectrum}, integrals with respect to the velocity space should be calculated. Numerically, the truncated but continuous velocity domain $\mathcal{D}$ needs to be represented by $M:=\left(M_1,M_2, M_3\right)$ discrete points $\bm v^{j'}$ and the integrals are approximated by certain quadrature rules, e.g. $\rho=\sum^{M}_{j'=1}f\left(\bm x,\bm v^{j'}\right)w^{j'}$ with $w^{j'}$ being the quadrature weight for the corresponding discretized velocity points $\bm v^{j'}$. The discrete velocities are not necessarily equidistant, especially for low-speed microflows with large Knudsen numbers, where the distribution function varies rapidly around $\bm v=0$ due to gas-wall interaction and nonuniform velocity points with refinement in this area is more efficient to capture the variation of $f$~\cite{PhysRevE.96.023309}. However, it should be emphasized that the FFT-based convolution could be efficiently employed only when the frequency space is uniformly discretized, though the number of frequency components can be smaller than that of velocity grid points due to the spectral accuracy of the FSM~\cite{wu2014}.

As a consequence, we need to approximate the distribution function at each discrete velocity point by solving $M_1\times M_2\times M_3\times K$ equations on each element $\Delta_i$ (take the ITR-MEAN scheme~\eqref{ITR1} as an example):
\begin{equation}
-\left(\nabla\varphi_s,\bm v^{j'}f^{j'}\right)_{\Delta_i}+\langle\varphi_s,\hat{\bm H}^{j'}\cdot\bm n\rangle_{\partial\Delta_i}+\left(\varphi_s,\bar{\nu} f^{j'}\right)_{\Delta_i}=\left(\varphi_s,\mathcal{C}^{j'}\right)_{\Delta_i}+\left(\varphi_s,\bar{\nu} f^{j'}\right)_{\Delta_i},
\label{dislocal}
\end{equation}
where $f^{j'}=f\left(\bm x,\bm v^{j'}\right)$, $\hat{\bm H}^{j'}=\hat{\bm H}\left(f^{j'},f^{j'}_\text{ext}\right)$, and $\mathcal{C}^{j'}=\mathcal{C}\left(f^{j'},f^{j'}_*\right)$ denote the corresponding variables at each discrete velocities. The resulting governing equations can be re-written into matrix form as:
\begin{equation}
\mathbf{A}^{i,j'}\mathbf{F}^{j'}_i+\mathbf{B}^{\text{ext},j'}=\mathbf{S}^{i,j'},
\label{linear}
\end{equation}
where $\mathbf{F}^{j'}_i=[F_1(\bm v^{j'}),\dots,F_r(\bm v^{j'}),\dots]^\mathrm{T}$ are the unknowns, i.e. the vector of degrees of freedom of $f^{j'}$ on $\Delta_i$. Other coefficient matrices are given in the Appendix. 

The strategy to solve the linear systems that are coupled through numerical fluxes over all spatial elements will be described in Sec.~\ref{Sweep}.

\subsection{Boundary condition} 

At the boundary of computational domain, to determine the flux for an element $\Delta_i$, the distribution function obtained from the exterior of the element, $f_\text{ext}$, is described by a given boundary condition $b^{j'}$. In this paper, the diffuse boundary condition is used at solid surface. Suppose the solid wall moves with a constant speed $\bm u_\text{w}$, and has a temperature $T_\text{w}$ that can either be a constant or vary along the wall, the distribution function for reflected molecules [i.e. when $\left(\bm v^{j'}-\bm u_\text{w}\right)\cdot\bm n_\text{w}\leq0$, $\bm n_\text{w}$ is the outward unit normal vector of the solid surface] is given by the equilibrium distribution:

\begin{equation}
b^{j'}=\frac{\rho_\text{w}}{\left(\pi T_\text{w}\right)^{3/2}}\exp\left(-\frac{|\bm v^{j'}-\bm u_\text{w}|^2}{T_\text{w}}\right),
\end{equation}
where, $\rho_\text{w}$ is defined by:

\begin{equation}
\sum_{\left(\bm v^{j'}-\bm u_\text{w}\right)\cdot\bm n_\text{w}<0}\left(\bm v^{j'}-\bm u_\text{w}\right)\cdot\bm n_\text{w}f^{j'}+\sum_{\left(\bm v^{j'}-\bm u_\text{w}\right)\cdot\bm n_\text{w}\leq0}\left(\bm v^{j'}-\bm u_\text{w}\right)\cdot\bm n_\text{w}b^{j'}=0
\end{equation}
such that the mass flux across wall is equal to zero. Implementation of other types of boundary conditions such as symmetry boundary, far-pressure inlet/outlet boundaries, and supersonic inlet/outlet boundaries can be found in Ref.~\cite{SU2015123}.

\section{Reduction of the Computational Complexity in DG Formalism}\label{ComputationComplex}

The major computational cost to solve the system~\eqref{linear} arises from two parts: 1) evaluating collision operator and 2) solving linear equations. In this section, we analysis the computational complexity for evaluation of the collision operator and left the one for solution of linear systems in Sec.~\ref{Sweep}. For simplification, we assume that equidistant discrete molecular velocities and frequencies are employed with $M=N$ and $N_1=N_2=N_3=\bar{N}$. Then, at each iterative step, equipped with the FFT-based convolution, the computational complexity is $O\left(K^2M_{\text{el}}M_\text{qua}^2\bar{N}^3\log\bar{N}+K^3M_\text{el}\bar{N}^3\right)$, in which the first term arises in the calculation of $\Xi_{p,r}$ and $\Lambda_{r}$ in Eq.~\eqref{DoFColl}, while the second term is for conducting the loops in Eqs.~\eqref{Direct2} and~\eqref{Direct1}.

Now, we propose an approach to reduce the cost in evaluating Boltzmann collision operator. In the following discussion, we will omit the index of discrete molecular velocities $j'$. The approach may be described heuristically in the following manner. If we choose the supporting polynomials as nodal shape functions:
\begin{equation}
\varphi_r\left(\bm x_p\right)=\begin{cases}
0,\quad\text{if}\ r\neq p,\\
1,\quad\text{if}\ r=p,
\end{cases}
\end{equation}
where $\bm x_p$ is the nodal points for interpolation, the degree of freedom $F_r$ is actually the nodal value of distribution function, say $f\left(\bm x_r\right)$. We assume that the distribution of $\mathcal{C}$ within an element might as well be estimated by the nodal approximation:
\begin{equation}
\mathcal{C}\simeq\mathcal{\tilde{C}}=\sum^{K}_{r=1}\varphi_r\left(\tilde{\Xi}_r-\Lambda_rF_r\right),
\label{Reduction}
\end{equation}
where $\tilde{\Xi}_r$ and $\Lambda_r$ are the nodal values of the collision gain term and collision frequency, respectively. The nodal value $\tilde{\Xi}_r$ is estimated from $F_r$ as $\tilde{\Xi}_r=\Xi_{r,r}$. As a result, the computational cost of $\left(\varphi_s,\mathcal{C}\right)_{\Delta_i}$ in Eqs.~\eqref{Direct2} and~\eqref{Direct1} is reduced to $O\left(KM_\text{el}M^2_\text{qua}\bar{N}^3\log\bar{N}+K^2M_\text{el}\bar{N}^3\right)$, that is, by $K$ times; this is considerable especially when high-order approximation polynomials are employed. For instance, nominally, 14 times less cost is expected when $k=4$ in 2D problems on triangular mesh.

It is interesting to note that, in the recent paper where an explicit DG Boltzmann solver has been developed, the singular value decomposition is proposed to reduce the computational cost~\cite{JAISWAL2018}. The singular value decomposition is pre-computed to the $K\times K$ matrix for $\left(\varphi_s,\varphi_p\varphi_r\right)$. Thus, the computational cost for the loops in Eqs.~\eqref{Direct2} and~\eqref{Direct1} can be reduced to $O\left(K^2M_\text{el}\bar{N}^3\right)$. However, the computational complexity for the collision operators by FSM [Eq.~\eqref{DoFColl}] remains unchanged and always consumes the majority of CPU time; thus this saving may not be in the order of magnitude.

Note that the introduced error of using Eq.~\eqref{Reduction} is proportional to:
\begin{equation}
\left(\varphi_s,|\mathcal{C}-\mathcal{\tilde{C}}|\right)_{\Delta_i}\varpropto\left(\varphi_s,|\bar{f}^{l}\bar{f}^{m}-\sum^{K}_{r=1}\varphi_r\bar{F}^{l}_r\bar{F}^{m}_r|\right)_{\Delta_i}+\left(\varphi_s,|\bar{f}^{j}\bar{f}^{j}-\sum^{K}_{r=1}\varphi_r\bar{F}^{j}_r\bar{F}^{j}_r|\right)_{\Delta_i},
\label{ERR}
\end{equation}
which is small when the variation of distribution function within a spatial element is not significant. In Sec.~\ref{Results}, we are going to valid this approximation numerically. The scheme with full calculation of collision terms~\eqref{Direct2} and~\eqref{Direct1} is labeled as `DG-FULL', while the one using reduced calculation~\eqref{Reduction} is labeled as `DG-RED'.

\section{Sweeping technique to solve the linear systems}\label{Sweep}

\begin{figure}[b]
	\begin{centering}
		\includegraphics[width=0.85\textwidth]{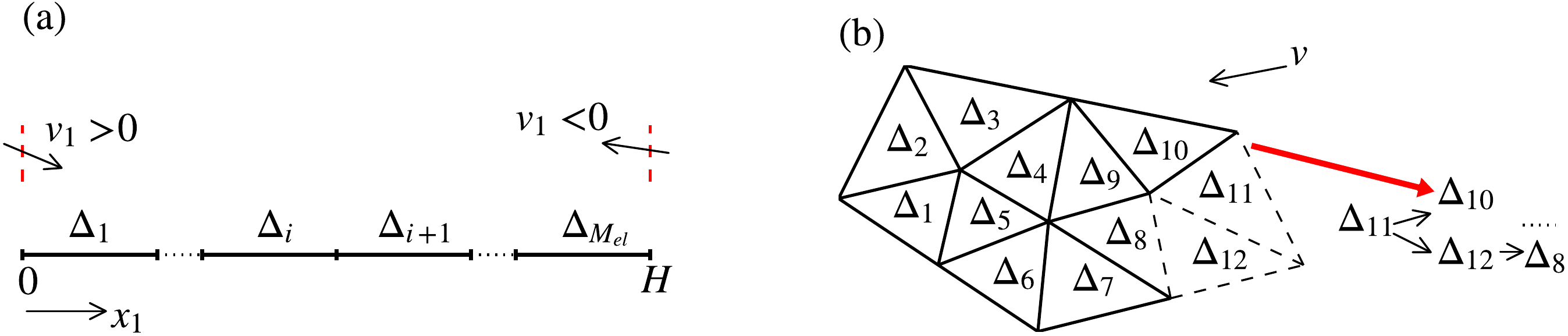}
		\par\end{centering}
	\caption{Schematic demonstration for determination of the spatial element ordering with respect to a given molecular velocity. (a) line mesh for 1D problem. (b) triangular mesh for 2D problem.}
	\label{Ordering}
\end{figure}

Now, we present the strategy to solve the linear systems resulting from the DG discretization. In the linear equations~\eqref{linear} on each spatial element $\Delta_i$, the unknown distribution function on neighboring element appears in $\mathbf{B}^{\text{ext},j'}$, the usual treatment in implicit DG is assembling the linear systems over all spatial elements and  solving a large sparse linear equation to determine the unknowns simultaneously. However, this requires huge memory and is prohibitively expensive in solving the Boltzmann equation, since we have to solve not one but a large number, e.g. several ten thousands, of large sparse linear systems. The matrix-free technique might be useful to improve the scheme~\cite{CRIVELLINI201181}. In this paper, a more intuitive and simpler strategy is adopted.

Due to the fact that the upwind flux is applied, it is important to notice that only the distribution function on neighboring elements in the upwind side appears in $\mathbf{B}^{\text{ext},j'}$. Thus, the solution of $f^{j'}$ on $\Delta_i$ can be obtained by solving the small linear system~\eqref{linear}, once $f_{\text{ext}}$ on the upwind side is known, or it is equipped with prescribed boundary conditions. Hence, starting from the element at the inflow boundary of computational domain, we can obtained the solution of $f^{j'}$ sequentially for all elements.

This sweeping technique, which requires no assembling large sparse linear system, relies on finding an ordering of the spatial elements, which is determined by the characteristic wind direction (that is, the direction of molecular velocity). The key to ensure feasible implementation of the sweeping technique in solving the Boltzmann equation is that, the discrete molecular velocity is fixed in the governing equations. Hence, we can find and store the spatial element ordering for each discrete velocity immediately after discretization and before the first iteration.

For a given discrete molecular velocity $\bm v^{j'}$, the topological ordering is easily found in 1D cases. As shown in Figure~\ref{Ordering}(a), a 1D computational domain $[0,H]$ is parallel to the $x_1$ axis. When $v^{j'}_1>0$, starting from the boundary at $x_1=0$, the spatial ordering is of ascending order in index $i$, while when $v^{j'}_1<0$ the spatial ordering has a descending order in $i$ starting from the boundary at $x_1=H$. For higher dimensional problems, we assume that the spatial grid is paved with convex elements and the element ordering is acyclic. A simple topological sorting algorithm is applied: gradually removing elements that have no incoming flux from elements left in the computational domain, placing them in the ordering, until no element remains. Figure~\ref{Ordering}(b) illustrates the schematic demonstration for the sorting procedure, where the ordering starts from the element $\Delta_{11}$ since it has only one inflow boundary located at the boundary of the computational domain. After removing $\Delta_{11}$, either element $\Delta_{10}$ or $\Delta_{12}$ will be put into the ordering, because there is no flux flowing from the elements left in the computational domain to these two elements. Note that the sequence of $\Delta_{11}$ and $\Delta_{12}$ in the ordering is interchangeable, since they do not share any common interface. The pseudo-code of the algorithm can be found in  Ref.~\cite{oatao14650} (Algorithm 3.2.2).

In Sec.~\ref{ComputationComplex}, we have mentioned that one of the majority consumptions in computational resources is to solve the linear systems. On the basis of the sweeping technique, if we use \textit{LU}-fabrication-based direct solver to solve the linear equations, the computational complexity is $O\left(2/3K^3M_\text{el}\bar{N}^3+2K^2M_\text{el}\bar{N}^3\right)$ since we have $M_\text{el}\bar{N}^3$ systems, and each has a coefficient matrix of rank $K$. Note that we have assumed that the number of discrete velocities in each direction is $\bar{N}$. In the ITR-MEAN scheme~\eqref{ITR1}, the complexity to solve linear equations can be reduced to $O\left(2K^2M_\text{el}\bar{N}^3\right)$ due to the fact that the coefficient matrix $\mathbf{A}^{i,j'}$ remains unchanged during all iterations and \textit{LU}-decomposition can be calculated and stored before the first iteration. The computational cost for \textit{LU}-decomposition is roughly $K/3$ times that for substitution in solving the linear equations, which becomes large as the grid density and/or the order of approximating polynomial increases. For example, when $k=4$ on triangular mesh, the computational complexity of \textit{LU}-decomposition is 4 times larger than that of substitution. Therefore, completing \textit{LU}-decomposition before iteration and only executing substitution during iteration can further save CPU time. 

\section{Numerical Results and Discussions}\label{Results}

The DG method with $k$ up to 4 is applied to solve the Boltzmann equation with full collision operator. The convergence criterion for the iterative schemes described above is that the global relative residual in the flow property $\mathcal{Q}$ between two successive iteration steps:
\begin{equation}
R_{\mathcal{Q}}=\frac{|\int_{\Delta}\mathcal{Q}^{\left(t+1\right)}-\mathcal{Q}^{\left(t\right)}\mathrm{d}\bm x|}{|\int_{\Delta}\mathcal{Q}^{\left(t\right)}\mathrm{d}\bm x|},
\end{equation}
is less than a threshold value $\epsilon$.

The following tests are performed in double precision on a workstation with Intel Xeon-E5-2680 processors and 132 GB RAM. During iteration, we call the routines in Intel Math Kernel Library (MKL) to conduct \textit{LU}-fabrication and solve linear equations. For the calculation of collision kernel $\beta\left(l,m\right)$, the trapezoidal rule is applied and we set $M_\text{qua}=5$ in Eq.~\eqref{GL_kernel} that is adequate to maintain the spectral accuracy of the FSM~\cite{WU2013}. Due to the fact that we only consider 1D and 2D flows, symmetry of the distribution function in the third ($v_3$) direction allows us to reduce the computational cost of Eq.~\eqref{GL_kernel} by half, that is, $\theta$ can be limited to the range of $[0, \pi/2]$; more details can be found in Ref.~\cite{wu2014}.

\subsection{1D normal shock wave}

\begin{table}[t]
\caption{Flow properties across normal shock waves. }

\centering{}%
\begin{tabular}{ccccc}
\hline
 & \multicolumn{2}{c}{$Ma=2.05$} &  \multicolumn{2}{c}{$Ma=9.0$}\tabularnewline
\hline
 & upstream & downstream & upstream & downstream \tabularnewline
\hline
$T$ & 1.0 & 2.144 & 1.0 & 26.185 \tabularnewline
$\rho$ & 1.0 & 2.334 & 1.0 & 3.857 \tabularnewline
$u_1$ & 1.871 & 0.802 & 8.216 & 2.130\tabularnewline
\hline
\end{tabular}
\label{SWcondition}
\end{table}

We first simulate the normal shock wave problem to assess the proposed method for the steady-state solution of the Boltzmann equation. Due to the absence of boundary effects, the flow is ideal to test the accuracy of DG discretization for streaming and the FSM approximation for the Boltzmann collision operator in capturing highly non-equilibrium, especially to validate the scheme with reduced DG calculation as described in Sec.~\ref{ComputationComplex}. The argon gas is considered with Mach numbers $Ma=2.05$ and $Ma=9.0$. We use the same parameters as those in Alsmeyer's experiments~\cite{alsmeyer1976}: the upstream density $\rho_0=1.067\times10^{-4}\ \text{kg/m}^3$ and temperature $T_0=300\ \text{K}$, corresponding to the mean free path and collision frequency of hard sphere molecules as $\bar{\lambda}=1.098\times10^{-3}\ \text{m}$ and $\bar{\nu}=3.633\times10^5~\text{s}^{-1}$, respectively. For all the DG results, the length scale is normalized with $H=\bar{\lambda}$ resulting in $Kn=5\pi/16$. The 1D computational domains $\Delta$ in the $x_1$ direction are $[-20,20]$ and $[-30,30]$ for $Ma=2.05$ and $Ma=9.0$ cases, respectively, which are partitioned by line elements with uniform length. The dimensionless up/downstream conditions normalized by the upstream properties are listed in Table~\ref{SWcondition}. Initially, the domains $x_1\leq0$ and $x_1>0$ are setup by the equilibrium distributions at upstream and downstream conditions, respectively. The implicit iteration scheme~\eqref{ITR2} with local collision frequency. i.e. ITR-LOC is applied. Iteration is terminated when $\max\{R_T,R_{n}, R_{|u_1|}\}<10^{-5}$. When $Ma=2.05$, the truncated velocity domain $[-8,8]^3$ is divided into $32^3$ uniform points, while when $Ma=9.0$, the velocity domain $[-30,30]^3$ is divided into $96\times64\times64$ uniform points. The same number of uniform frequencies are used for approximation of the collision operator.

\begin{figure}
	\begin{centering}
		\includegraphics[width=0.8\textwidth]{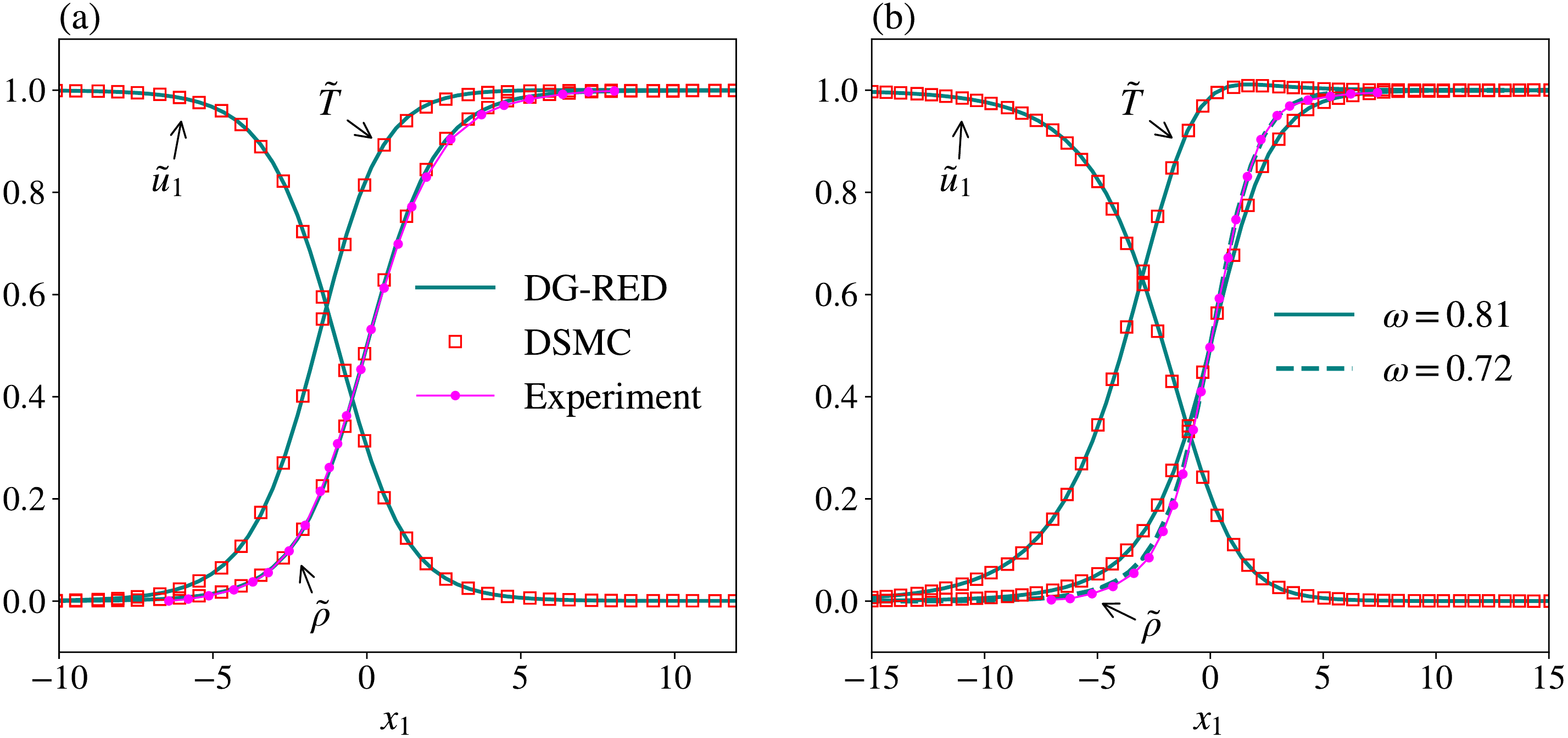}
		\par\end{centering}
	\caption{Profiles of normalized flow velocity $\tilde{u}_1=\frac{u_1-u_{1,\text{R}}}{u_{1,\text{L}}-u_{1,\text{R}}}$, density $\tilde{\rho}=\frac{\rho-\rho_\text{L}}{\rho_\text{R}-\rho_\text{L}}$ and temperature $\tilde{T}=\frac{T-T_\text{L}}{T_\text{R}-T_\text{L}}$ for normal shock wave of argon gas at (a) $Ma=2.05$ and (b) $Ma=9.0$. The subscripts `L' and `R' denote the properties in upstream and downstream, respectively. The DG-RED solutions are obtained with $k=4$, $M_\text{el}=16$. The ITR-LOC scheme~\eqref{ITR2} is applied for implicit iteration.}
	\label{SW1}
\end{figure}

Numerical tests show that by using the sweeping technique, the implicit DG method is stable without any limiter in solving the 1D normal shock structure. Figure~\ref{SW1} illustrates the DG results of normalized flow velocity, density and temperature, compared with the DSMC results and experimental data~\cite{alsmeyer1976}. The DSMC results presented here are computed using the code developed and verified in~\cite{Tang2014}. In order to ensure accuracy of the DSMC method, the cell sizes and time steps are set to be $\sim0.13\bar{\lambda}$ and $\sim0.12/\bar{\nu}$, respectively. The average number of molecules per spatial cell is about 50. About 30,000 iterations are needed to reach the steady-state solutions. To obtained smooth results, macroscopic flow properties are sampled over another 100,000 steps. For comparison, the viscosity index in both methods are set as $\omega=0.81$. The DG results are obtained using 4\textsuperscript{th}-order approximating polynomial on 16 elements, which  agree well with those of DSMC simulation (the profiles from the DG-FULL scheme are not shown, since they overlap with the ones of DG-RED). We also compare the DG solutions for density with the experimental data. For $Ma=2.05$, the agreement is good, although slight discrepancy can be observed in the downstream side of the shock wave. For $Ma=9.0$ where the non-equilibrium effect is strong, the DG solutions agrees well with the DSMC ones. However, disagreement between the DG (DSMC) solution and experimental one enlarges, where the variation of density is steeper in experiment. Actually, the profiles in high Mach number flow are more sensitive with respect to the value of viscosity index $\omega$. The works in~\cite{XU20107747,Valentini2009} suggest that to set $\omega$ being around 0.7, the Boltzmann solver or DSMC can produce result closed to the experimental one. Hence, we include the DG-RED solution with $\omega=0.72$ (dash lines in Figure~\ref{SW1}(b)), and obtain an improved agreement.      

\begin{figure}
	\begin{centering}
		\includegraphics[width=0.95\textwidth]{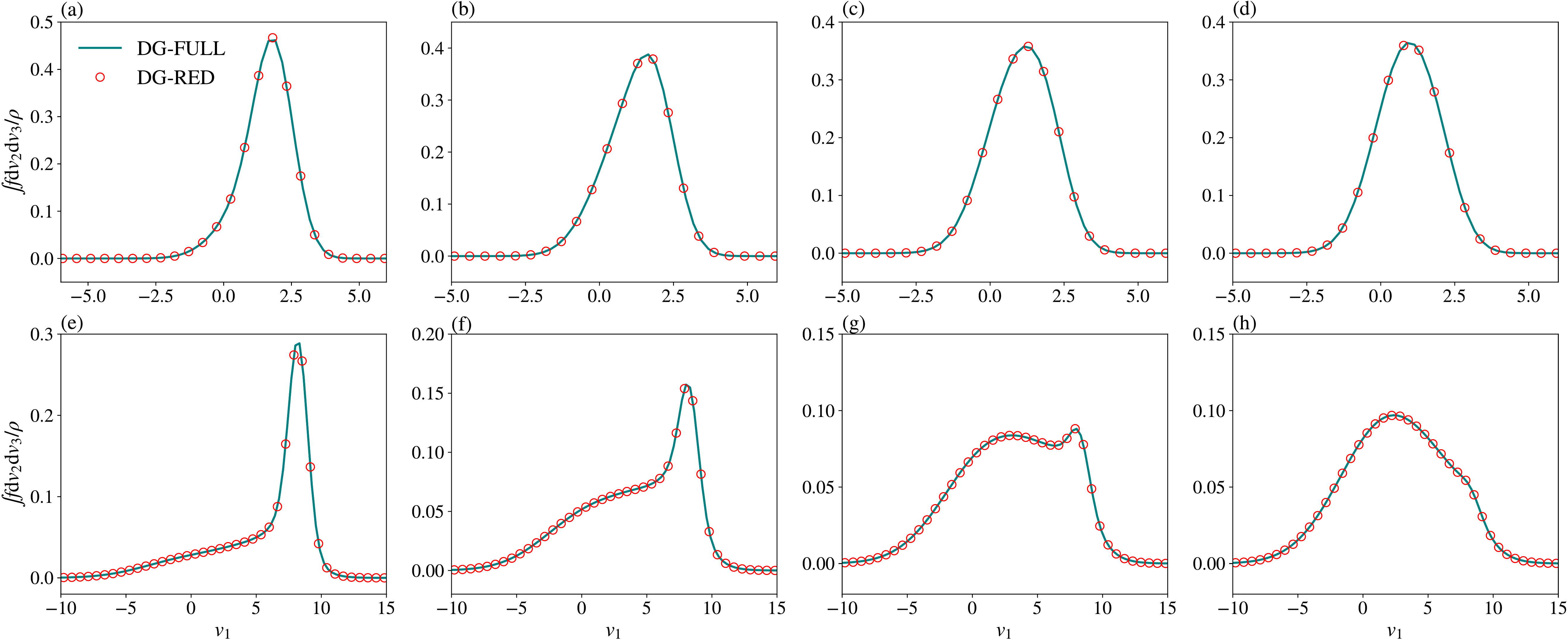}
		\par\end{centering}
	\caption{Comparison of the marginal distribution functions $\int f\mathrm{d}v_2\mathrm{d}v_3/\rho$ from the DG-RED and the DG-FULL schemes: first row is ones for $Ma=2.05$ presented at (a) $\rho=1.197$, (b) $\rho=1.468$, (c) $\rho=1.766$ and (d) $\rho=1.991$; second row is ones for $Ma=9.0$ presented at (e) $\rho=1.423$, (f) $\rho=1.943$, (g) $\rho=2.543$ and (h) $\rho=3.123$. For the DG-FULL, the molecular velocity domains $[-8,8]^3$ and $[-30,30]^3$ are uniformly divided into $64\times32\times32$ and $192\times64\times64$ points for $Ma=2.05$ and 9.0, respectively. The ITR-LOC scheme~\eqref{ITR2} is applied for implicit iteration.}
	\label{SW2}
\end{figure}

To further validate the DG-RED scheme, we compare the marginal distribution functions $\int f\mathrm{d}v_2\mathrm{d}v_3$ at different locations of the shock wave with those obtained using the DG-FULL scheme. The profiles are plotted in Figure~\ref{SW2}. To ensure accuracy of the DG-FULL results, we have doubled the number of discrete velocity and frequency points in the longitudinal direction. It is demonstrated that in low Mach number flow, the distribution functions are closed to the corresponding equilibrium (Gaussian) distribution. As Mach number increases, the distribution functions within the shock wave structure greatly deviate from the equilibrium states. The comparison shows that, even for highly non-equilibrium flow, the DG-RED scheme can produce correct solutions, so that the numerical error brought by the reduced calculation of collision operator is negligible.

Another important property of a shock wave with $Ma>\sqrt{9/5}$ in a monatomic gas is the overshoot of temperature associated with the longitudinal component of thermal velocities, $T_x$, which could be larger than the gas temperature behind the front of shock due to the non-equilibrium in translational energies of longitudinal and transversal directions. The analytical form of $T_x$ is related to the density $\rho$ as~\cite{Yen1966}:

\begin{equation}
T_{x,\text{an}}=\frac{1}{3}\left[\frac{\left(5Ma^2+3\right)}{\rho}-5\left(\frac{Ma}{\rho}\right)^2\right].
\end{equation}
Based on $T_x$, we compare the convergence behavior of DG-RED and DG-FULL schemes with respect to various orders of approximating polynomials $k$ and numbers of spatial elements $M_\text{el}$. The relative $L_2$ error of $T_x$ that is evaluated as
\begin{equation}
\mathcal{E}=\frac{\int_{\Delta}\left(T_x-T_{x,\text{an}}\right)^2\mathrm{d}x_1}{\int_{\Delta} T^2_{x,\text{an}}\mathrm{d}x_1},
\end{equation} 
the number of iteration steps and the total CPU time are listed in Table~\ref{SW}.

All tests are done on single processor, and the internal parallelism for MKL functions is not activated. It is shown that for each $k$, as the number of elements increases, errors of $T_x$ gradually converges to 0.016\% and 0.036\% for Mach numbers of 2.05 and 9.0, respectively. The higher order approximating polynomials, the fewer elements needed to obtain the converged results. The numbers of iterative steps to reach the steady-state solutions also converge to fixed values of around 201 and 225 for Mach numbers of 2.05 and 9.0, respectively. Therefore, compared to the lower-order scheme, the higher-order discretization consumes less CPU time to obtain solution with the same order of accuracy. For example, for $Ma=2.05$, the DG-FULL scheme with $k=4$ cost about 30\% less CPU time to produce solution with $\mathcal{E}=0.016\%$ on the mesh of 16 segments, compared to the one with $k=3$ that obtains the same accurate result on 32 segments.

It is found that the DG-RED scheme can preserve these convergence properties. That is, by using the same order of approximating polynomials on the same mesh, the DG-RED and DG-FULL require the same number of iterative step to obtain solutions of the same order of accuracy. However, the DG-RED can significantly save the computational cost in terms of CPU time. The higher degree of approximating polynomials, the more the saving. For example, for $Ma=9.0$, to obtained solution of $\mathcal{E}=0.036\%$, both the schemes need 64, 32 and 16 spatial elements for $k=2$, 3 and 4, respectively, and the CPU time consumed by the DG-RED is about 50\%, 41\%, and 36\% of that by the DG-FULL. 


\begin{sidewaystable}
\caption{Comparisons between the DG-FULL (with FULL calculation of $\left(\varphi_s,\mathcal{C}\right)_{\Delta_i}$) and the DG-RED (with reduced calculation of $\left(\varphi_s,\mathcal{C}\right)_{\Delta_i}$) in terms of the relative $L_2$ error ($\mathcal{E}$) of longitudinal temperature $T_x$ (compared with the analytical result), the number of iterations (Itr denotes the number of iteration steps to reach the convergence criterion $\max\{R_T,R_{\rho},R_{|u_1|}\}<10^{-5}$), and the CPU time $t_\text{c}$. Normal shock wave is considered. The ITR-LOC scheme~\eqref{ITR2} is applied for implicit iteration.}

\centering{}%
\begin{tabular}{cccccccccccccccccc}
\hline
\multirow{3}{*}{k} & \multicolumn{8}{c}{$Ma=2.05$} &  & \multicolumn{8}{c}{$Ma=9.0$}\tabularnewline
\cline{2-9} \cline{11-18} 
 & \multicolumn{4}{c}{DG-FULL} &  & \multicolumn{3}{c}{DG-RED} &  & \multicolumn{4}{c}{DG-FULL} &  & \multicolumn{3}{c}{DG-RED}\tabularnewline
\cline{2-5} \cline{7-9} \cline{11-14} \cline{16-18}
 & $M_{\text{el}}$ & $\mathcal{E}\times10^{-2}$ & Itr & $t_{c}$, {[}h{]} &  & $\mathcal{E}\times10^{-2}$ & Itr & $t_{c}$, {[}h{]} &  & $M_{\text{el}}$ & $\mathcal{E}\times10^{-2}$ & Itr & $t_{c}$, {[}h{]} &  & $\mathcal{E}\times10^{-2}$ & Itr & $t_{c}$, {[}h{]}\tabularnewline
\hline
\multirow{4}{*}{1} & 4 & 14.656 & 208* & 0.02 &  & 7.734 & 204** & 0.01 &  & 8 & 12.91 & 201 & 0.64 & & 5.814 & 278 &  0.61 \tabularnewline
 & 8 & 3.450 & 194 & 0.03 &  & 2.159 & 219 & 0.02 &  & 16  & 1.668 & 222 & 1.32 & & 1.553 & 252 & 1.08 \tabularnewline
 & 16 & 0.643 & 199 & 0.06 &  & 0.622 & 207 & 0.03 &  & 32 & 0.396 & 225 & 2.57 & & 0.381 & 238 & 1.96 \tabularnewline
 & 32 & 0.158 & 201 & 0.11 &  & 0.152 & 203 & 0.06 &  & 64 & 0.088 & 225 & 5.19 & & 0.087 & 229 & 3.72 \tabularnewline
\hline
\multirow{4}{*}{2} & 4 & 3.343 & 189 & 0.03 &  & 1.629 & 203 & 0.01 &  & 8 & 0.824 & 225 & 1.39 & & 0.361 & 231 & 0.75 \tabularnewline
 & 8 & 0.202 & 199 & 0.06 &  & 0.187 & 198 & 0.02 &  & 16  & 0.239 & 225 & 2.73 & & 0.247 & 226 & 1.49 \tabularnewline
 & 16 & 0.097 & 200 & 0.13 &  & 0.097 & 200 & 0.04 &  & 32 & 0.049 & 225 & 5.75 & & 0.049 & 225 & 2.86 \tabularnewline
 & 32 & 0.020 & 201 & 0.24 & & 0.020 & 201 & 0.08 &  & 64  & 0.036 & 225 & 11.24 & & 0.036 & 225 & 5.52 \tabularnewline
\hline
\multirow{4}{*}{3} & 4 & 0.221 & 190 & 0.05 &  & 0.350 & 201 & 0.02 &  & 8 & 0.550 & 225 & 2.58 &  & 0.595 & 235 & 1.06 \tabularnewline
 & 8 & 0.216 & 198 & 0.09 & & 0.219 & 199 & 0.03 &    & 16 & 0.066 & 225 & 5.06 &  & 0.065 & 226 & 1.98 \tabularnewline
 & 16 & 0.022 & 200 & 0.21 & & 0.022 & 200 & 0.06 &   & 32 & 0.036 & 225 & 9.38 &  & 0.036 & 225 & 3.88 \tabularnewline
 & 32 & 0.016 & 201 & 0.41 &  & 0.016 & 201 & 0.11 &  & 64 & 0.036 & 225 & 16.22 &  & 0.036 & 225 & 7.49 \tabularnewline
\hline
\multirow{4}{*}{4} & 4 & 0.557 & 189 & 0.09 &  & 0.584 & 188 & 0.02 &  & 8 & 0.236 & 225 & 3.89 &  & 0.233 & 226 & 1.30 \tabularnewline
 & 8 & 0.060 & 198 & 0.15 &  & 0.061 & 198 & 0.04 &   & 16 & 0.037 & 225 & 7.35 & & 0.037 & 225 & 2.53 \tabularnewline
 & 16 & 0.016 & 200 & 0.30 &  & 0.016 & 201 & 0.07 &  & 32 & 0.036 & 225 & 13.15 & & 0.036 & 225 & 4.85 \tabularnewline
 & 32 & 0.016 & 202 & 0.59 &  & 0.016 & 201 & 0.14 &  & 64 & 0.036 & 225 & 25.85 & & 0.036 & 225 & 9.55 \tabularnewline
\hline
\end{tabular}
{\footnotesize Due to round-off errors, *this case only converged to residual of about $2\times10^{-5}$ and **this case only converged to residual of about $3\times10^{-5}$.}
\label{SW}
\end{sidewaystable}

\subsection{Hypersonic flow past a square cylinder}

\begin{figure}[t]
	\begin{centering}
		\begin{subfigure}
		\centering
		\includegraphics[width=0.42\textwidth]{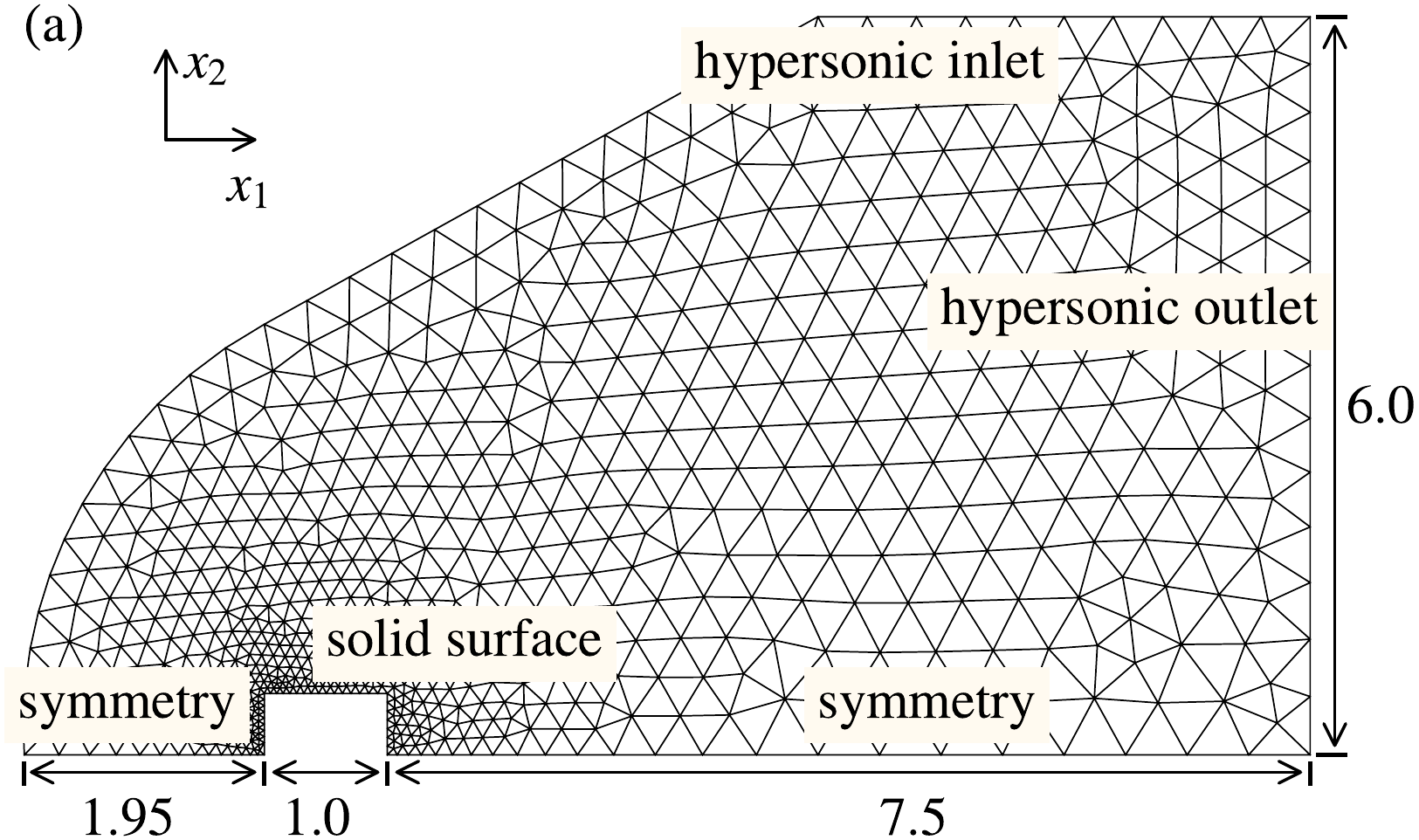}
		\end{subfigure}
		\begin{subfigure}
		\centering
		\includegraphics[width=0.42\textwidth]{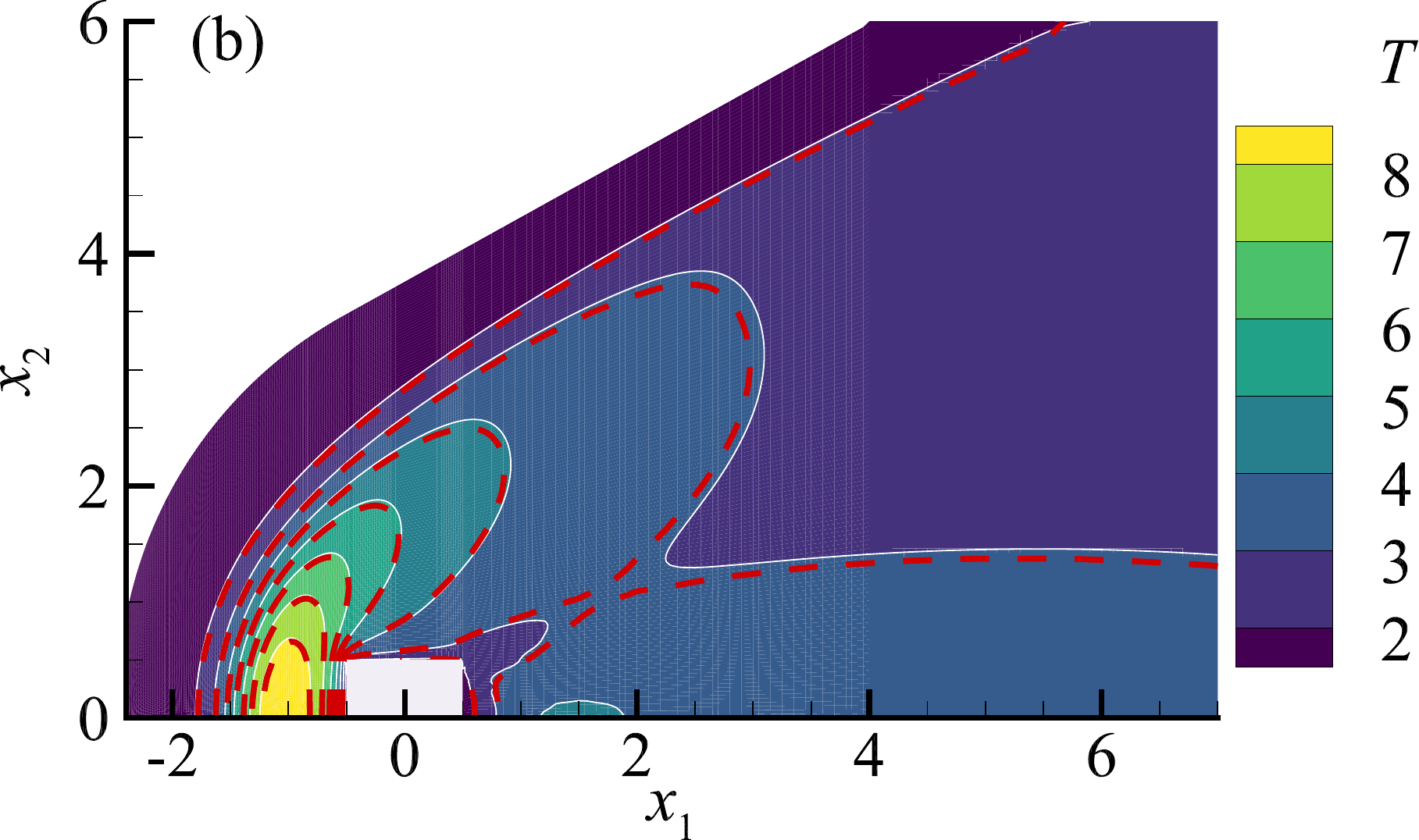}
		\end{subfigure}
		\\
		\begin{subfigure}
		\centering
		\includegraphics[width=0.42\textwidth]{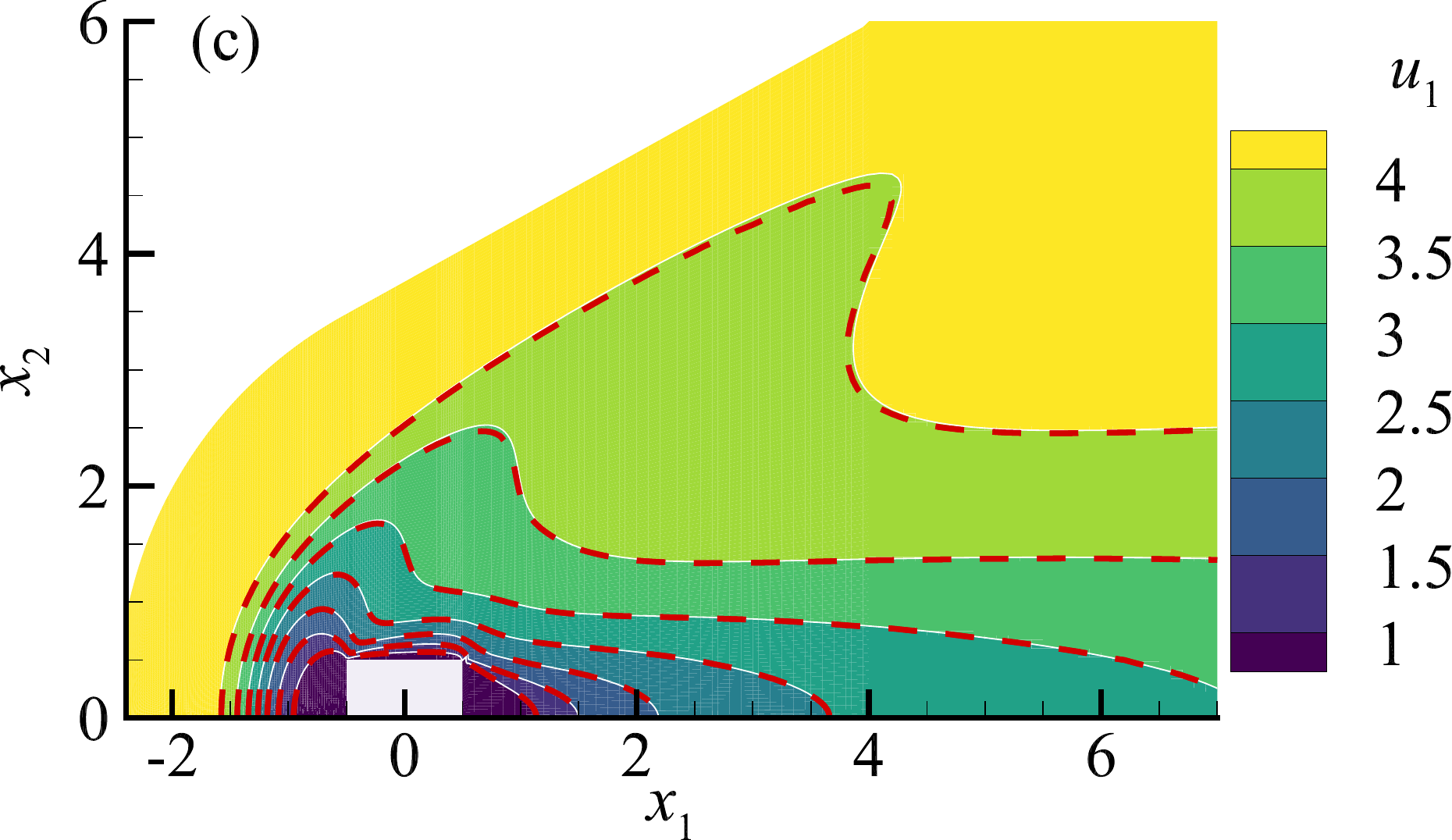}
		\end{subfigure}
		\begin{subfigure}
		\centering
		\includegraphics[width=0.42\textwidth]{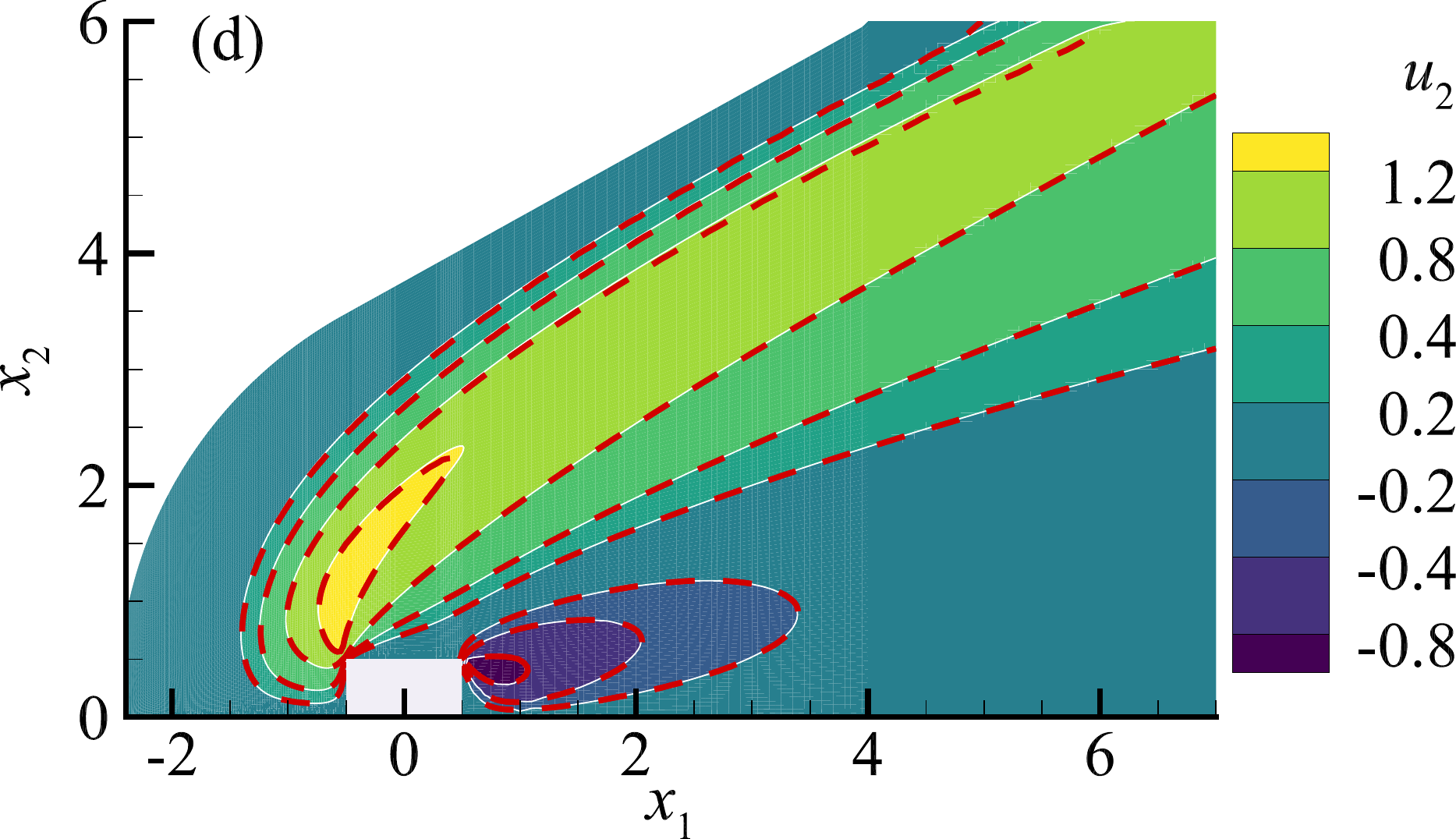}
		\end{subfigure}
		\par\end{centering}
	\caption{Hypersonic flow of $Ma=5$ and $Kn=0.13$ past a square cylinder. (a) Schematic for computational domain, boundary condition and unstructured triangular mesh. (b) Temperature contours. (c) Horizontal velocity contours. (d) Vertical velocity contours. The white solid contour lines with background illustrate the solution from the DG-RED of $k=4$, where the molecular velocity domain $[-13,13]^3$ is discretized by $48\times48\times48$ uniform grid points. The red dashed contour lines are the DSMC results in Ref.~\cite{CHEN2017119}.}
	\label{SquareField}
\end{figure}

Now we consider a 2D high-speed flow. the DG-RED scheme of $k=4$ is applied to compute hypersonic flow past a square cylinder having a dimension of $1\times1$ and a constant wall temperature of $T_\text{w}=1.0$. The free stream has dimensionless temperature and density of $T_0=1.0$ and $\rho_0=1.0$. The Mach number and Knudsen number in the free stream are set as 5.0 and 0.13, respectively. As shown in Figure~\ref{SquareField}(a), the argon gas of viscosity index $\omega=0.81$ moves from left to right along the $x_1$ direction. The computational domain is chosen with extension up to 1.95, 7.5 and 5.5 away from the cylinder in the upwind, downstream and $x_2$ direction, respectively. Due to symmetry, only half of the flow field is considered. The boundary conditions and triangular mesh are also illustrated in Figure~\ref{SquareField}(a). Besides the full-diffuse solid surfaces, the lower boundaries parallel to $x_1$ is symmetric boundaries, while other boundaries are set as hypersonic inlet/outlet boundaries where the distribution function is the equilibrium distribution at free-stream condition. 1490 unstructured triangles are employed to discretize the computational domain, with refinement near the solid surfaces. The truncated molecular velocity space $[-13,13]^3$ are discretized by $48\times48\times48$ uniform points, and the same number of uniform frequencies are used for evaluation of collision terms. The flow field is initialized by the free-stream condition, and the ITR-LOC scheme~\eqref{ITR2} is applied which costs about 346 steps to reach the convergence criterion of $\max\{R_T,R_{\rho},R_{|\bm u|}\}<5\times10^{-5}$. The test is run on 28 processors using OpenMP for parallelism and consumes about 24.6 hours of wall time. 

At the very beginning of iteration, strong discontinuity appears in the upwind side of the square cylinder due intense stagnation effect of gas flow, and the DG scheme can generate spurious oscillation which may make the approximated distribution functions negative. As a consequence, the loss term will become the gain term and the iteration will lead to unphysical blowup solutions. To tackle this problem, instead of using any nonlinear limiters as one usually does, we take absolute values to the negative degrees of freedom after solving the linear systems at each iterative step. Numerical test shows that this simple treatment does not destroy accuracy of the DG discretization but does guarantee its stability.

\begin{figure}[t]
	\begin{centering}
		\begin{subfigure}
		\centering
		\includegraphics[angle=90,width=0.32\textwidth]{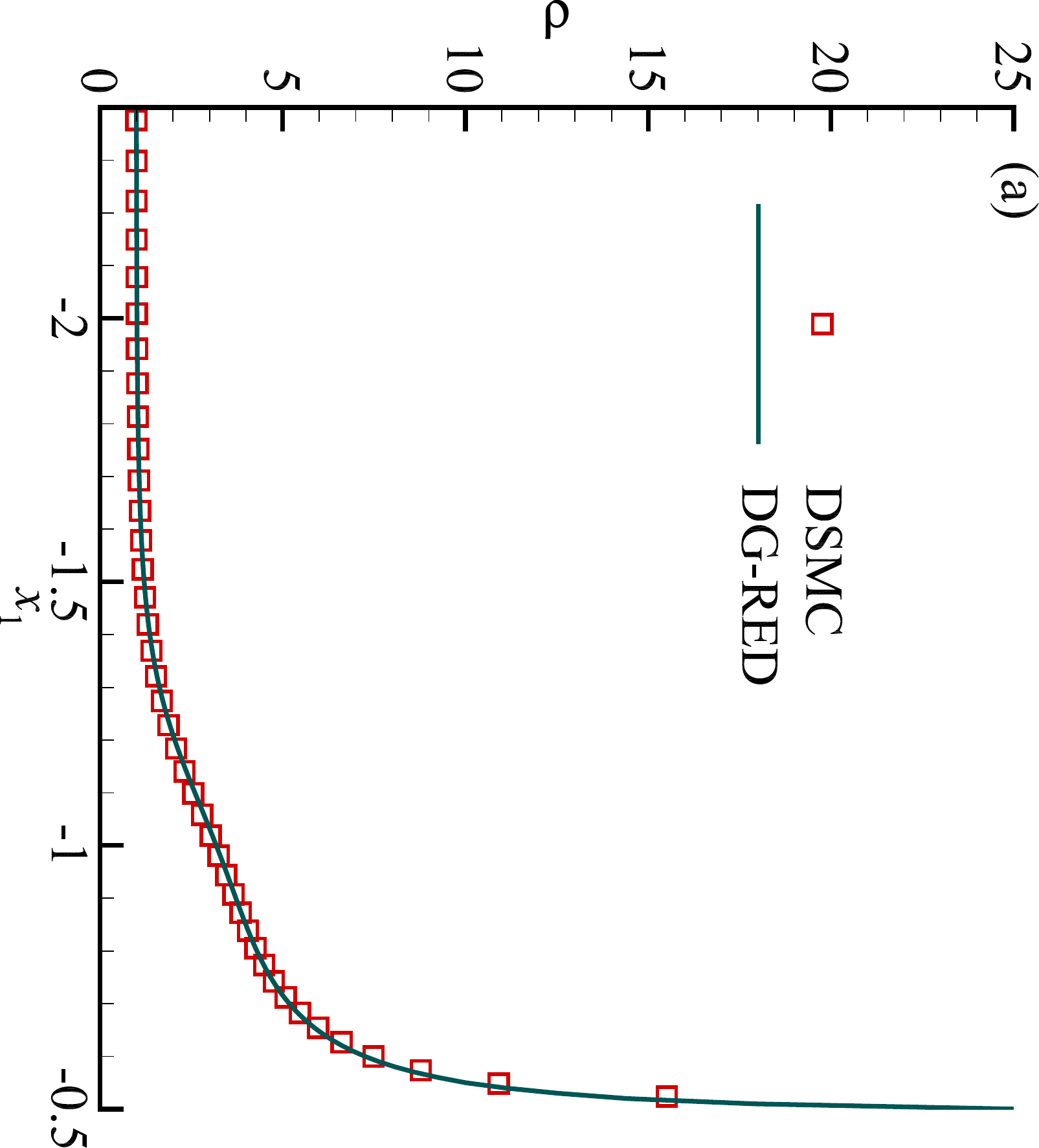}
		\end{subfigure}
		\begin{subfigure}
		\centering
		\includegraphics[width=0.32\textwidth]{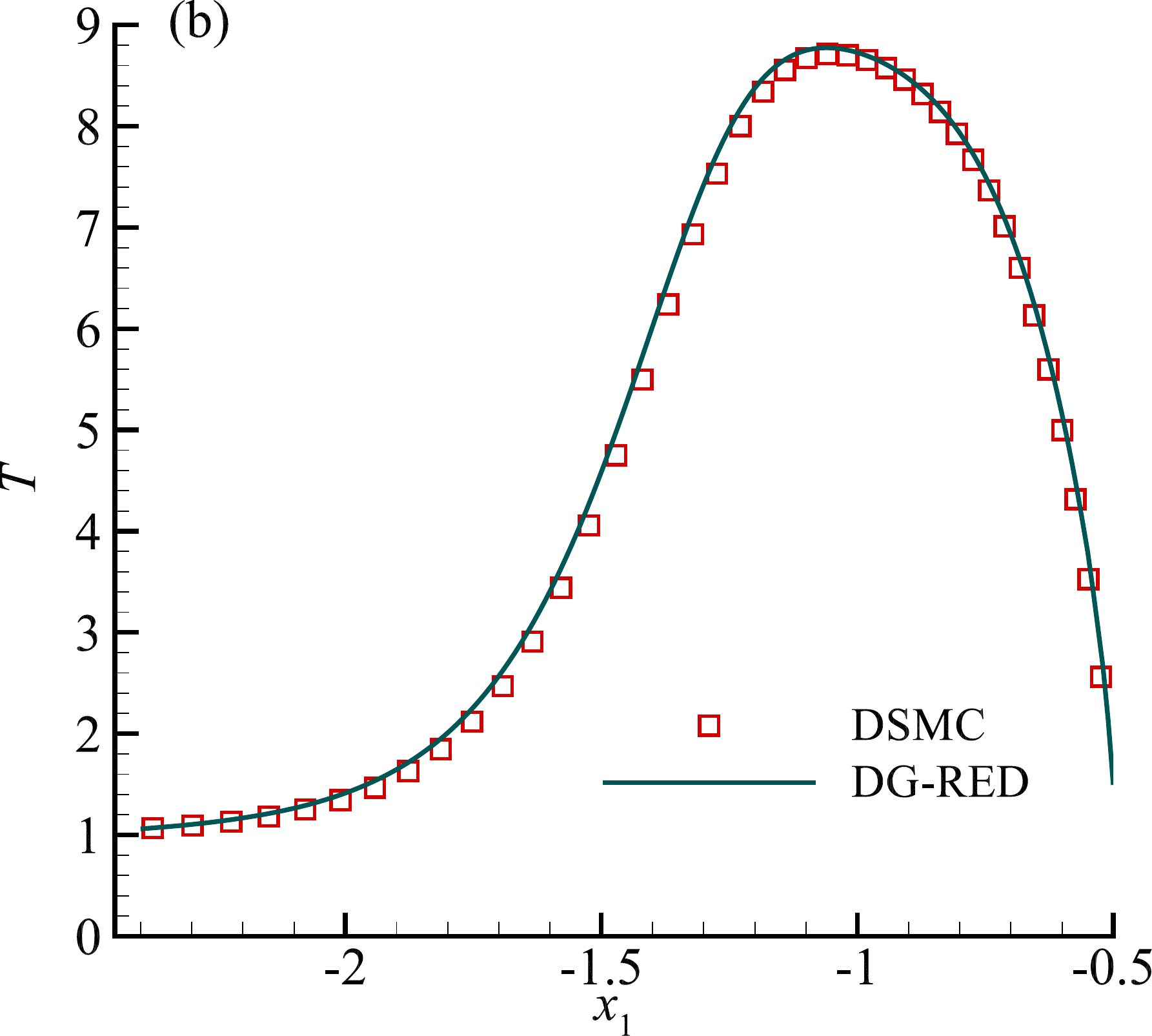}
		\end{subfigure}
		\begin{subfigure}
		\centering
		\includegraphics[width=0.32\textwidth]{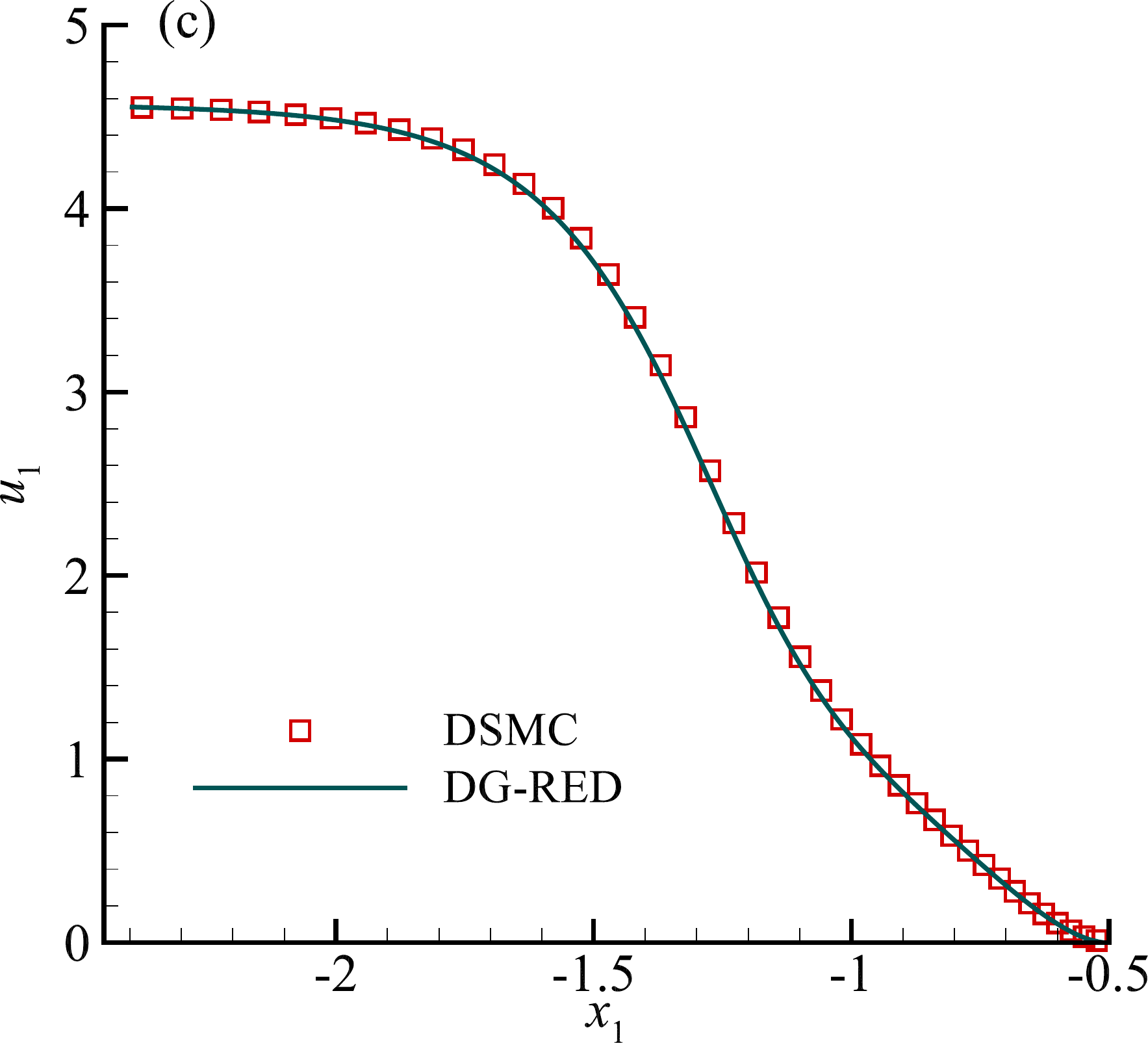}
		\end{subfigure}
		\par\end{centering}
	\caption{Hypersonic flow of $Ma=5$ and $Kn=0.13$ past a square cylinder. Profiles of (a) density (b) temperature and (c) horizontal velocity along the symmetric line in front of the stagnation point. Solid lines are the solutions from the DG-RED of $k=4$, where the molecular velocity domain $[-13,13]^3$ is discretized by $48\times48\times48$ uniform grid points. Symbols are the DSMC results in Ref.~\cite{CHEN2017119}.}
	\label{SquareStagnation}
\end{figure}

Contours of temperature, horizontal velocity and vertical velocity are illustrated in Figure~\ref{SquareField}(b)-(d). The white lines with background are the DG-RED solution, while the red dashed contour lines are the DSMC results in Ref.~\cite{CHEN2017119}. Note that the Knudsen number in~\cite{CHEN2017119} is $2\left(7-2\omega\right)\left(5-2\omega\right)/15\pi$ times the unconfined Knudsen number in this paper. Comparison between the DG-RED solutions and DSMC ones on the distributions of density, temperature and horizontal velocity along the symmetric line in the front of the stagnation point are shown in Figure~\ref{SquareStagnation}. It is found that due to the stagnation effect from the static cylinder to the gas flow, the flow density increases about 25 times within 10 (free-streaming) mean free paths when approaching to the cylinder, and the bulk horizontal velocity drops to zero. Since the isothermal wall condition is applied, the flow temperature first increases to its maximum value of 8.7 at about 5 mean free paths away from the stagnation point and then decreases to 1.45 at the solid wall.   

\begin{figure}
	\begin{centering}
		\includegraphics[width=0.7\textwidth]{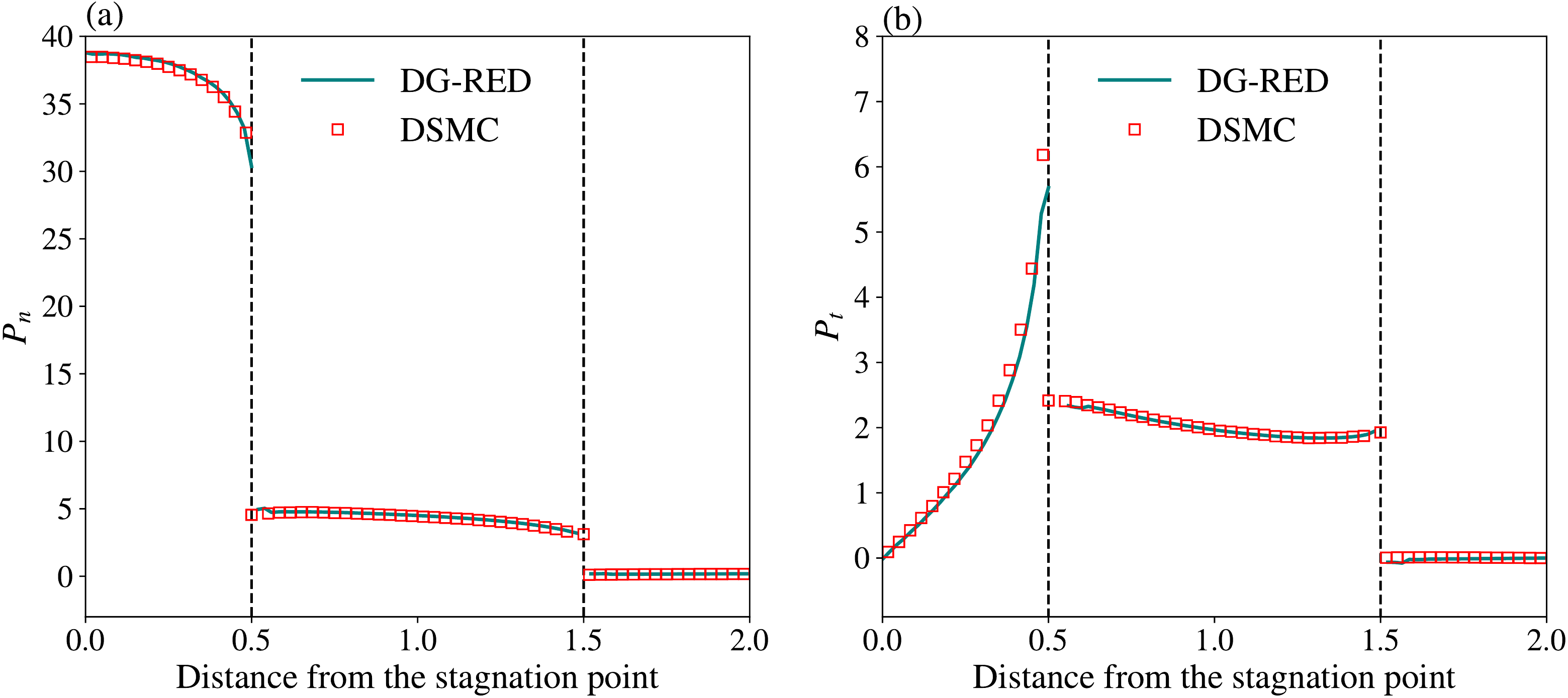}
		\par\end{centering}
	\caption{Hypersonic flow of $Ma=5$ and $Kn=0.13$ past a square cylinder. The distributions of (a) normal stress and (b) shear stress along the surface of cylinder. The horizontal axis represents the distance along the surface of the square, starting from the stagnation point in a counter-clockwise direction. Solid lines are the DG-RED solutions and symbols are the DSMC results in Ref.~\cite{CHEN2017119}. }
	\label{SquareSurface}
\end{figure}

Figure~\ref{SquareSurface} illustrates the distributions of normal stress $P_\text{n}$ and shear stress $P_\text{t}$ along the surfaces of the square cylinder, where $P_\text{n}=\bm n_\text{w}\cdot\bm P\cdot\bm n_\text{w}$ and $P_\text{t}=\bm t_\text{w}\cdot\bm P\cdot\bm t_\text{w}$ with $\bm n_\text{w}$ and $\bm t_\text{w}$ denoting the outward unite normal vector and tangential vector of the solid surface, respectively. The largest $P_\text{n}$ is at the surface in the upwind side where the normal momentum flux is large, while the shear stress gradually increases along that surface as the bulk vertical velocity increases. Both the $P_\text{n}$ and $P_\text{t}$ vary slightly along the top surface and the lateral surface in the weak. Figures~\ref{SquareField} to~\ref{SquareSurface} demonstrate the good agreement between the DG and DSMC results. 

\subsection{2D lid-driven cavity flow}

By comparing with the DSMC results, a 2D low-speed flow in a square cavity driven by the top lid is used to compare performances of the DG Boltzmann solvers and a Boltzmann solver using the second-order FDM to approximate derivatives in the spatial space~\cite{Wu2015}. The wall temperature is set as the reference temperature $T_0=273$ K. The velocity of the driven lid is 50 m/s. The flow gas is argon with a viscosity index of 0.81. The gas flow is initialized to be rest at $T_0$ with $Kn=1$, where the characteristic length $H$ is chosen to be the side length of the square cavity. The computational configuration for DSMC can be found in~\cite{John2010}.

\begin{figure}[t]
	\begin{centering}
		\begin{subfigure}
			\centering
			\includegraphics[width=0.239\textwidth]{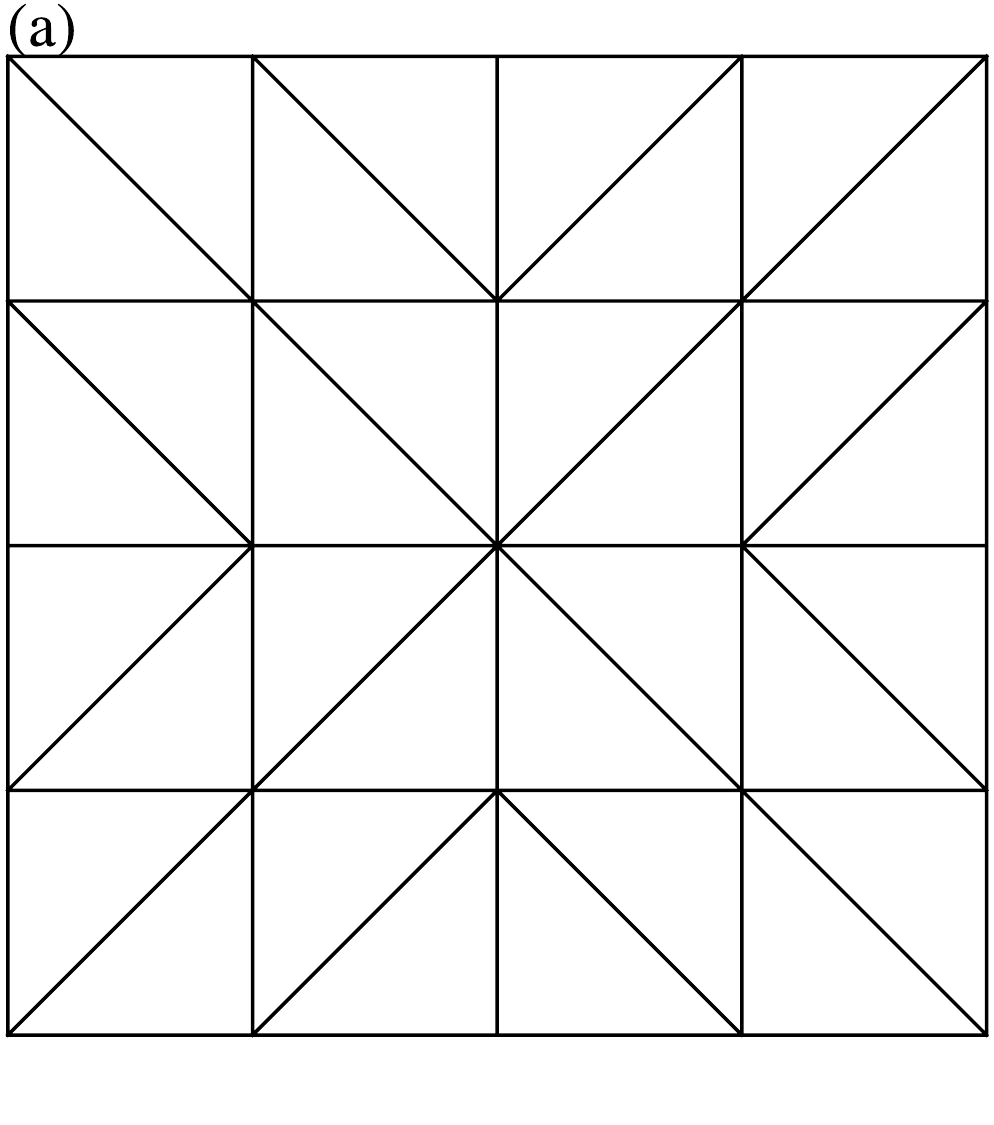}
		\end{subfigure}
		\begin{subfigure}
			\centering
			\includegraphics[width=0.34\textwidth]{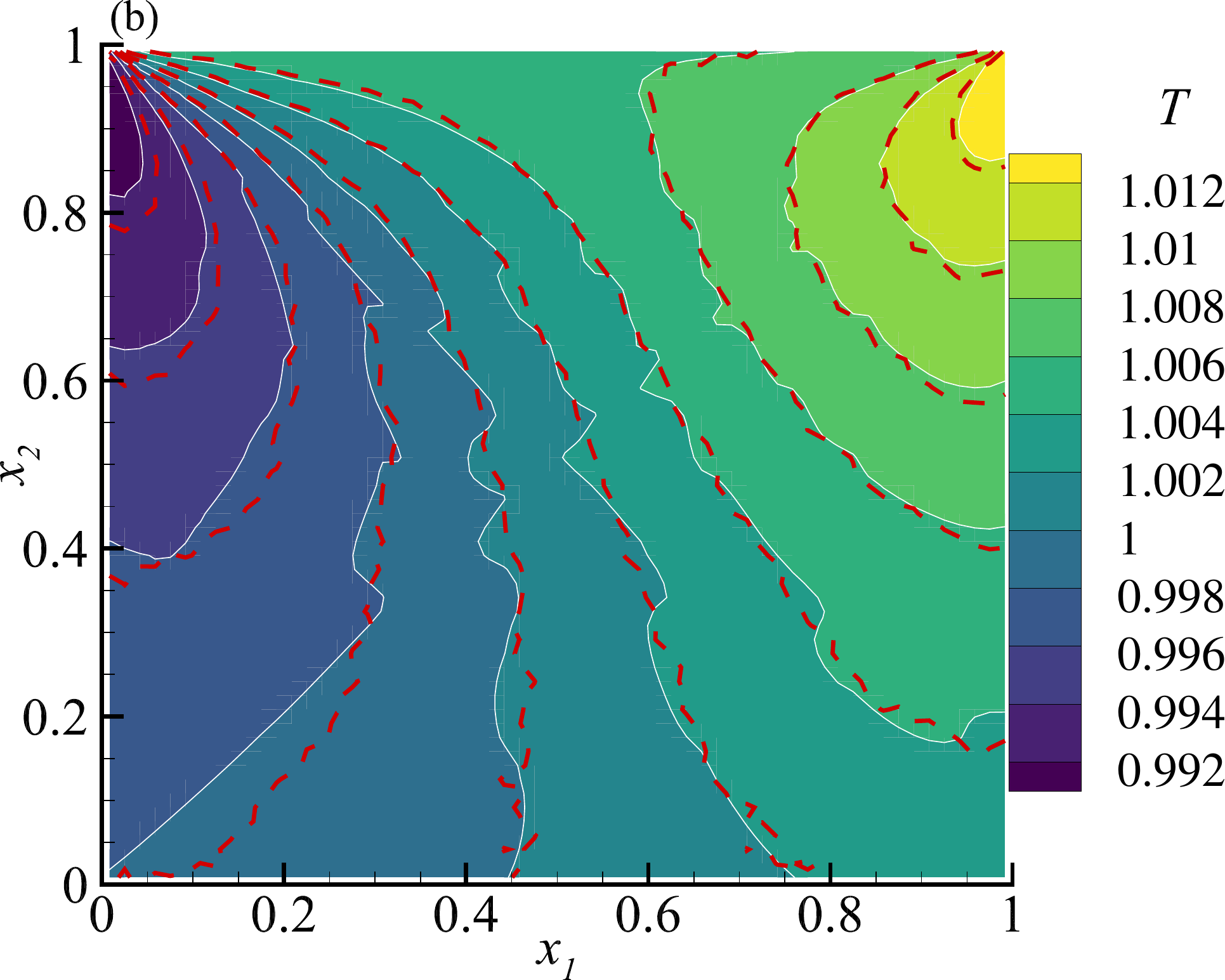}
		\end{subfigure}
		\begin{subfigure}
			\centering
			\includegraphics[width=0.34\textwidth]{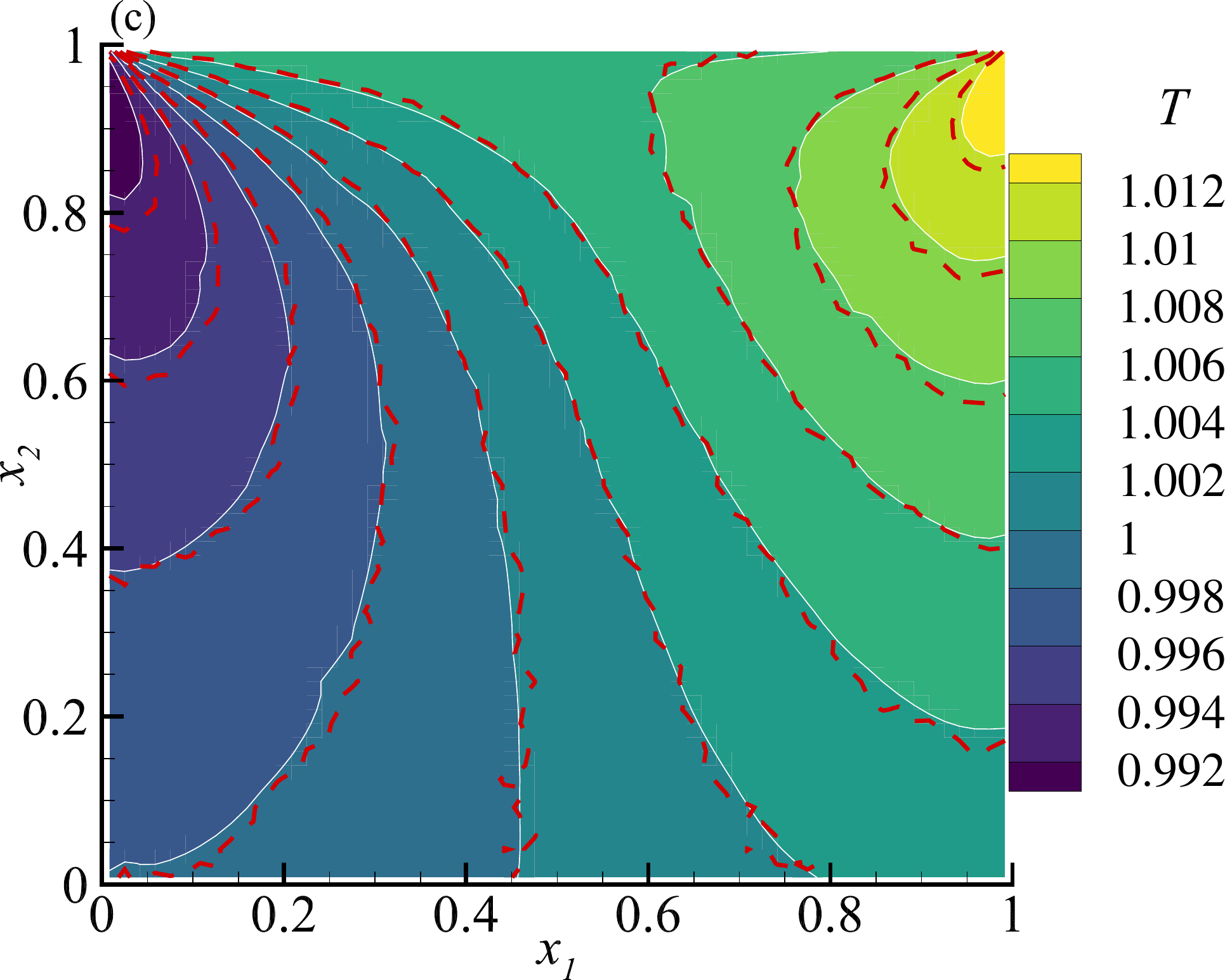}
		\end{subfigure}
	\par\end{centering}
	\caption{Comparison of the DG-RED and the DSMC on square cavity flow at $Kn=1.0$ driven by moving lid with speed of $U_0=0.148$. The DG-RED solutions are obtain with $k=4$ on 72 uniform triangles. (a) typical triangular mesh; (b) and (c): temperature contours when the molecular velocity domain $[-6,6]^3$ is discretized by $36\times36\times24$ and $108\times108\times24$ grid points in the DG-RED, respectively. White solid lines with background indicate solution of the DG-RED, while red dashed lines are the DSMC result. The ITR-LOC scheme~\eqref{ITR2} is applied for implicit iteration.}
	\label{Cavity1}
\end{figure}

For deterministic solutions, the truncated molecular domain is selected as $[-6,6]^3$. The DG and FDM solvers utilize the same FSM to evaluate collision terms in frequency domain, which is discretized with $32\times32\times24$ equidistant frequencies. For discretization in the molecular velocity, non-uniform points are used for $v_1$ and $v_2$, while uniform discrete velocities are used in the third direction. The non-uniform discretization with refinement around $v_{1(2)}=0$ is efficient to calculate low-speed flows especially at large Knudsen numbers, where the distribution function changes rapidly within a narrow area around the origin in the $v_1$ and $v_2$ directions~\cite{PhysRevE.96.023309}. For spatial discretization, uniform triangular mesh is used in the DG method, as shown in Figure~\ref{Cavity1}(a), while the FDM uses equidistant grid points in the $x_1$ and $x_2$ directions. Determination on the numbers of spatial elements and discrete velocities is a trivial task. General speaking, flows with small values of $Kn$ need relatively large number of spatial elements to ensure that the artificial diffusion is much smaller than the physical viscosity that is small in near-continuum flows, while highly rarefied flows require a large number of discrete velocities to resolve significant variations and/or discontinuities in the distribution function. Moreover, the spatial and velocity grids have `contrary' effects, where finite discretization of the velocity space tends to capture discontinuities, whereas limited spatial discretization tends to smooth flow field due to artificial diffusion. Incompatible spatial and velocity grids can lead to emergence of the so called `ray effect', which causes deterministic solution oscillating around its mean value~\cite{Chai1993,COELHO2002231}. To overcome this shortcoming, the velocity grid should be fine enough so that error induced by the ray effect is small, which can be compensated by the error of numerical diffusion~\cite{COELHO2002231}.

Temperature contours from the DG-RED for $k=4$ and $M_\text{el}=72$ (highly resolved in the spatial space) are compared with the DSMC results in Figure~\ref{Cavity1}. Results in  Figure ~\ref{Cavity1}(b) and (c) are obtained with $36\times36\times24$ and $108\times108\times24$ velocities, respectively. It is observed that relative coarser velocity grid produces temperature contour with violent fluctuations, and refinement in the velocity discretization can largely improve the accuracy. Besides, DG solver with higher order of approximating polynomial is more likely to suffer the ray effect. This is mainly due to the fact that, compared to lower-order scheme, higher-order scheme can obtain more accurate result on same spatial grid so that the numerical diffusion is relatively smaller which can not smear the ray effect.

\begin{figure}[t]
	\begin{centering}
		\includegraphics[width=0.85\textwidth]{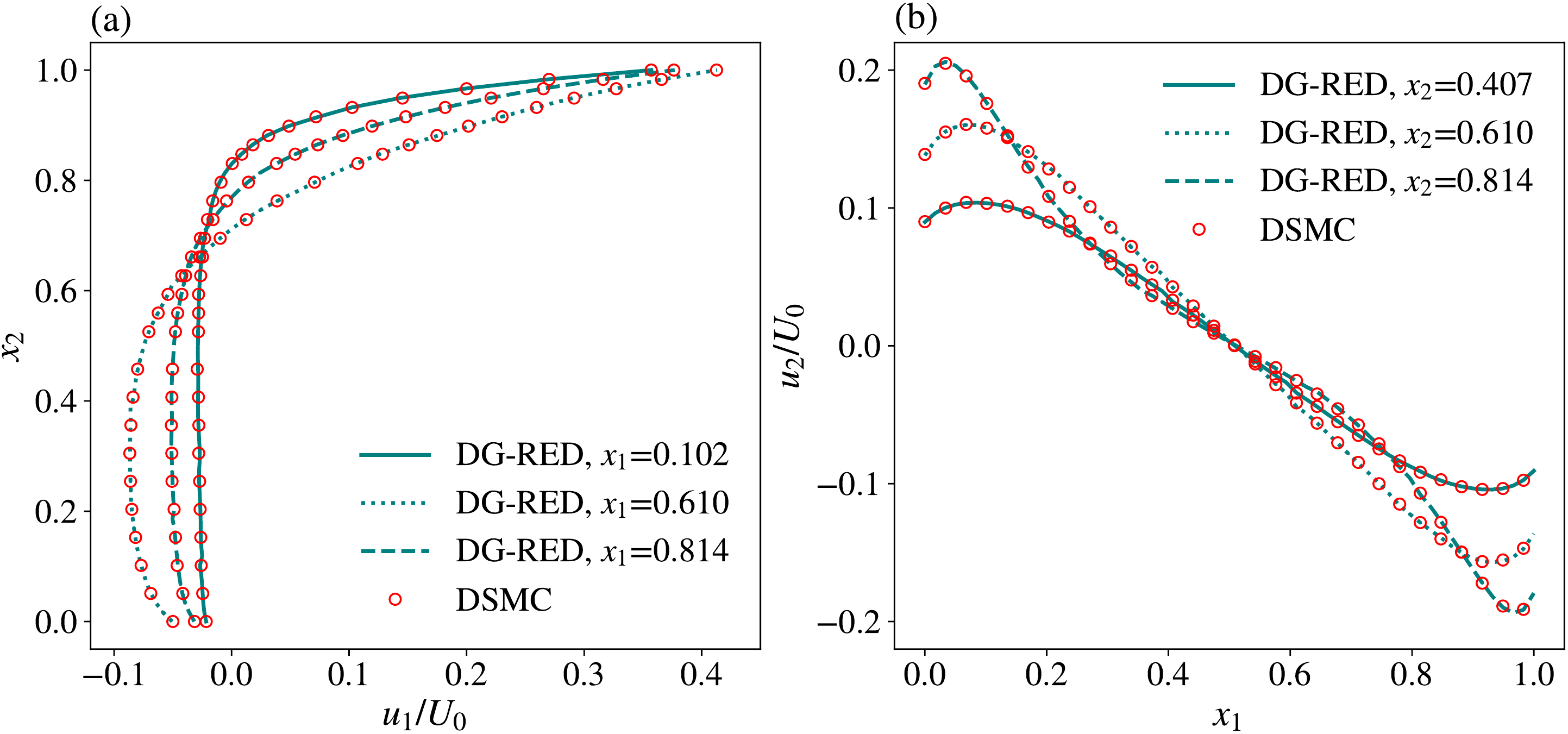}
		\par\end{centering}
	\caption{Comparison of the DG-RED and the DSMC on square cavity flow at $Kn=1.0$ driven by moving lid with speed of $U_0=0.148$. (a) normalized horizontal flow velocity $u_1/U_0$ along vertical lines at different locations; (b) normalized vertical flow velocity $u_2/U_0$ along horizontal lines at different locations. The DG-RED solutions are obtained with $k=4$ and 72 triangles. The molecular velocity domain $[-6,6]^3$ is discretized by $108\times108\times24$ grid points. The ITR-LOC scheme~\eqref{ITR2} is applied for implicit iteration.}
	\label{Cavity2}
\end{figure}

Further comparison on the results of DG-RED and DSMC are illustrated in Figure~\ref{Cavity2} in terms of horizontal (vertical) flow velocity along selected vertical (horizontal) lines. The DG-RED results possess good agreement with those of the DSMC.

In Table~\ref{CavityDG}, we list the relative $L_2$ error of velocity magnitude $|\bm u|$, the number of iterations to reach the convergence criterion $\max\{R_T,R_n,R_{\bm u}\}<10^{-5}$, as well as the total CPU time cost for the DG-FULL with the ITR-LOC, the DG-RED with the ITR-LOC and the DG-RED with the ITR-MEAN. In the ITR-MEAN iterative scheme~\eqref{ITR1}, the mean collision frequency is set as $\bar{\nu}=1.4$ for this specific flow. For all cases, the molecular velocity domain $[-6,6]^3$ is discretized by $72\times72\times24$ grid points. The errors are calculated in reference to the DSMC results, which are obtained at $60\times60$ equidistant points $\bm x_p$ in the computational domain. The errors are evaluated as
\begin{equation}\label{DSMC_lid}
\mathcal{E}=\sqrt{\sum\left(|\bm u\left(\bm x_p\right)|_\text{DG}-|\bm u\left(\bm x_p\right)|_\text{DSMC}\right)^2/\sum|\bm u\left(\bm x_p\right)|^2_\text{DSMC}},
\end{equation} 
where the DG solution at any point $\bm x_p$ can be easily obtained through polynomial approximation. All tests are done on single processor. It is shown that, for each order of approximating polynomials, the three schemes can produce solution with the same accuracy on the same spatial mesh. The iterative scheme using local collision frequency can obtain the steady-state solution within 21 steps, no matter which DG calculation (DG-FULL or DG-RED) is applied. Thus, due to the reduction of computational complexity in calculation of the Boltzmann collision operator, the DG-RED cost less CPU time than the DG-FULL. Equipped with the chosen mean collision frequency, the ITR-MEAN iterative scheme~\eqref{ITR1} uses 17 steps to reach the steady-state solution. Since it does not require \textit{LU}-decomposition during iterations, scheme combining the DG-RED and the ITR-MEAN can further reduce the computational cost. For example, to obtain solution of error in velocity magnitude equal to 0.014 with $k=4$ and $M_\text{el}=18$, the DG-RED plus the ITR-MEAN costs about 50\% and 92\% less CPU time than that of the DG-RED with the ITR-LOC and the DG-FULL with the ITR-LOC, respectively.

\begin{table}[t]
\caption{Comparisons between the DG-FULL with the ITR-LOC iteration, and the DG-RED with the ITR-LOC as well as the ITR-MEAN in the lid-driven square cavity flow with $Kn=1.0$, in terms of the relative $L_2$ error $\mathcal{E}$~\eqref{DSMC_lid}, the number of iterations (Itr denotes the number of iteration steps to reach the convergence criterion $\max\{R_T,R_{\rho},R_{|\bm u|}\}<10^{-5}$), and the CPU time $t_\text{c}$. The molecular velocity domain $[-6,6]^3$ is discretized by $72\times72\times24$ grid points.}

\centering{}
\begin{tabular}{ccccccccccccc}
\hline
\multirow{2}{*}{$k$} & \multirow{2}{*}{$M_\text{el}$} & \multicolumn{3}{c}{DG-FULL + ITR-LOC} & & \multicolumn{3}{c}{DG-RED + ITR-LOC} & & \multicolumn{3}{c}{DG-RED + ITR-MEAN}\tabularnewline
\cline{3-5} \cline{7-9} \cline{11-13}
 & & $\mathcal{E}$ & Itr & $t_\text{c}, {[}h{]}$ & & $\mathcal{E}$ & Itr & $t_\text{c}, {[}h{]}$ & & $\mathcal{E}$ & Itr & $t_\text{c}, {[}h{]}$\tabularnewline
 \hline
 \multirow{4}{*}{1} 
  & 32 & 0.102 & 21 & 0.044 & & 0.102 & 21 & 0.024 & & 0.102 & 17 & 0.018 \tabularnewline
  & 50 & 0.080 & 21 & 0.068 & & 0.080 & 21 & 0.038 & & 0.080 & 17 & 0.029 \tabularnewline
  & 72 & 0.065 & 21 & 0.100 & & 0.065 & 21 & 0.056 & & 0.065 & 17 & 0.042 \tabularnewline
  & 98 & 0.054 & 21 & 0.258 & & 0.054 & 21 & 0.077 & & 0.054 & 17 & 0.059 \tabularnewline
 \hline
\multirow{4}{*}{2} 
  & 32 & 0.039 & 21 & 0.154 & & 0.039 & 21 & 0.058 & & 0.039 & 17 & 0.043 \tabularnewline
  & 50 & 0.029 & 21 & 0.244 & & 0.029 & 21 & 0.080 & & 0.029 & 17 & 0.059 \tabularnewline
  & 72 & 0.023 & 21 & 0.457 & & 0.023 & 21 & 0.125 & & 0.023 & 17 & 0.088 \tabularnewline
  & 98 & 0.019 & 21 & 0.895 & & 0.019 & 21 & 0.173 & & 0.019 & 17 & 0.126 \tabularnewline
 \hline
 \multirow{4}{*}{3} 
  & 18 & 0.025 & 21 & 0.222 & & 0.025 & 21 & 0.056 & & 0.025 & 17 & 0.033 \tabularnewline
  & 32 & 0.019 & 21 & 0.551 & & 0.019 & 21 & 0.120 & & 0.019 & 17 & 0.082 \tabularnewline
  & 50 & 0.014 & 21 & 0.950 & & 0.014 & 21 & 0.180 & & 0.014 & 17 & 0.117 \tabularnewline
  & 72 & 0.012 & 21 & 1.078 & & 0.012 & 21 & 0.272 & & 0.012 & 17 & 0.190 \tabularnewline
 \hline
 \multirow{4}{*}{4} 
  & 8  & 0.024 & 21 & 0.219 & & 0.024 & 21 & 0.044 & & 0.024 & 17 & 0.023 \tabularnewline
  & 18 & 0.014 & 21 & 0.696 & & 0.014 & 21 & 0.111 & & 0.014 & 17 & 0.057 \tabularnewline
  & 32 & 0.011 & 21 & 1.321 & & 0.011 & 21 & 0.234 & & 0.011 & 17 & 0.143 \tabularnewline
  & 50 & 0.008 & 21 & 2.013 & & 0.008 & 21 & 0.349 & & 0.008 & 17 & 0.224 \tabularnewline
 \hline
\end{tabular}
\label{CavityDG}
\end{table}

We also list the error of velocity magnitude, the number of iterations and the CPU time for the FDM in Table~\ref{CavityFDM}. Uniformly distributed points are employed to discretize the spatial space. Thus, the computational domain is partitioned by rectangular elements and flow properties are evaluated at the vertices of rectangles. To estimate the error of velocity magnitude, $\bm u\left(\bm x_p\right)$ may not associated to a discrete grid point, then it is obtained through linear interpolation using the four values at vertices of the grid cell that $\bm x_p$ locates in. The FDM solver also uses 21 steps to obtain steady-state solutions, since the ITR-LOC iterative scheme~\eqref{ITR2} is employed. For comparison of the DG and the FDM, we find that the DG discritization is more efficient. For instance, the FDM predicts solution with error in $|\bm u|$ of 0.015 on the spatial grid with $71\times71$ grid points, while the DG scheme achieves solution with the same order of accuracy on 50 and 18 triangles for $k=3$ and 4, respectively. However, the DG method with $k=3$ and full calculation in collision terms cost more CPU time than the FDM. This is because, although the computational complex for the Boltzmann collision operator in the DG-FULL with $k=3$ and $M_\text{el}=50$ ($\varpropto M_\text{el}K^2$) and in the FDM with $M_\text{p}=71\times71$ ($\varpropto{}M_\text{p}$) is similar,  the DG scheme requires additional time to solve linear equations. As a consequence, only the DG-RED scheme can preserve the efficiency of DG in terms of CPU time. Equipped with the ITR-LOC iteration~\eqref{ITR2}, to obtain solution with error in $|\bm u|$ of 0.015, the DG-RED solvers of $k=3$ and 4 are about 4 and 7 times faster than the FDM. The ITR-MEAN iteration~\eqref{ITR1} can further boost its efficiency, now the DG-RED solvers of $k=3$ and 4 can be 6 and 13 times faster than the FDM. Although higher-order FDM could achieve better efficiency, it needs much more computational effort since stencils involving large numbers of points are required. Also, it has difficulty to treat complex geometries. 

\begin{table}[t]
\caption{Performance of the FDM combining the ITR-LOC iteration~\eqref{ITR2} for solution of lid-driven square cavity flow at $Kn=1.0$. $M_\text{p}$ is the number of discrete points in the spatial space. $\mathcal{E}$ is the relative $L_2$ error of velocity magnitude $|\bm u|$ compared with the DSMC results. Itr denotes the number of iteration steps to reach the convergence criterion $\max\{R_T,R_{\rho},R_{|\bm u|}\}<10^{-5}$. $t_\text{c}$ is the total CPU time. The molecular velocity domain $[-6,6]^3$ is discretized by $72\times72\times24$ non-uniform grid points.}

\centering{}
\begin{tabular}{ccccccccc}
\hline
$M_\text{p}$ & $\mathcal{E}$ & Itr & $t_\text{c}, {[}h{]}$ & & $M_\text{p}$ & $\mathcal{E}$ & Itr & $t_\text{c}, {[}h{]}$\tabularnewline
 \hline
$31^2$ & 0.052 & 21* & 0.159 & & $61^2$ & 0.018 & 21 & 0.611 \tabularnewline
$41^2$ & 0.046 & 21 & 0.282 & & $71^2$ & 0.015 & 21 & 0.845 \tabularnewline
$51^2$ & 0.028 & 21 & 0.433 & & $81^2$ & 0.016 & 21 & 1.065\tabularnewline
 \hline
\end{tabular}

{\footnotesize{} *This case only converged to residual of about $1.2\times10^{-5}$ due to round-off errors.}
\label{CavityFDM}
\end{table}

\subsection{2D flow induced by a hot micro-beam in a rectangular chamber}

\begin{figure}[t]
	\begin{centering}
		\includegraphics[width=0.65\textwidth]{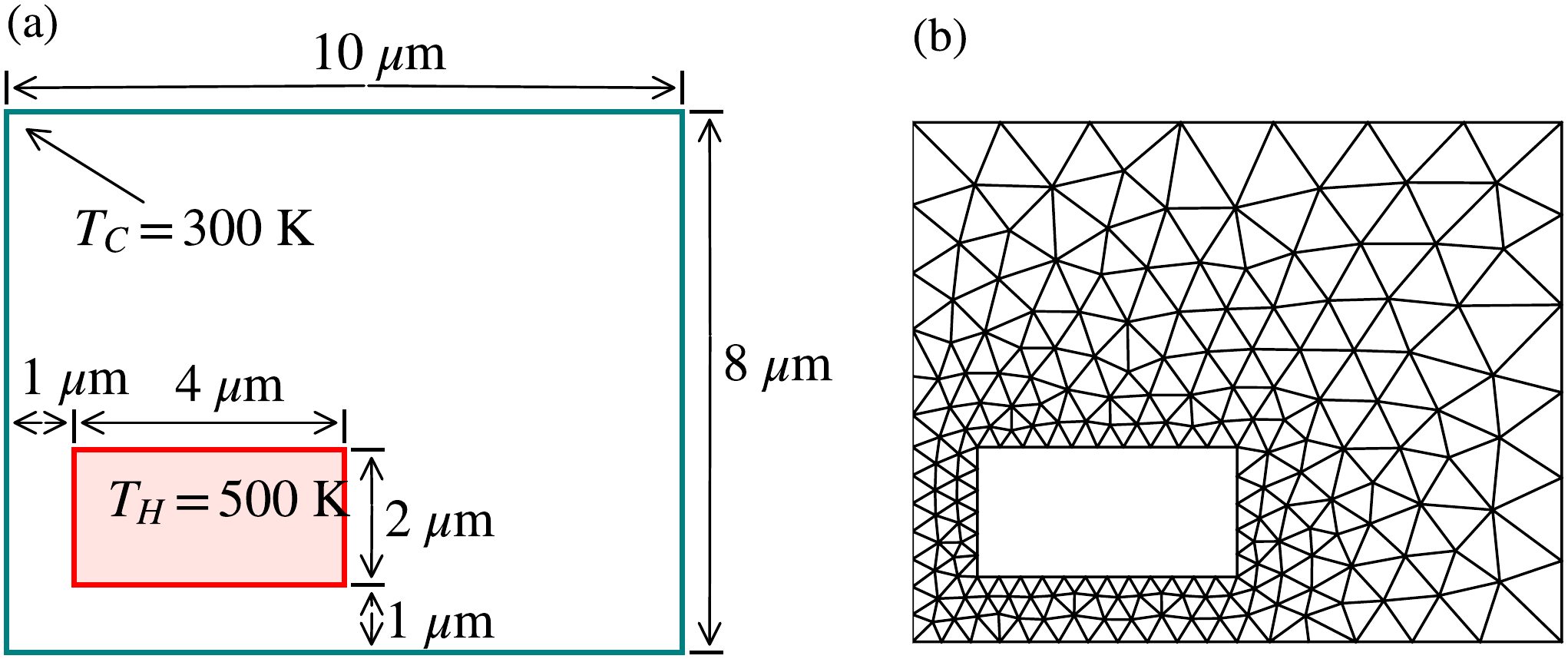}
		\par\end{centering}
	\caption{The micro gas flow around heated beam in a rectangular chamber. (a) geometry and (b) schematic of the triangular mesh.}
	\label{BeamG}
\end{figure}

We then consider the performance of the DG method in simulation of low-speed rarefied gas flow inside micro-channel.  As depicted in Fig.~\ref{BeamG}(a), we consider a 2D rarefied gas flow induced by a hot micro-beam with a thickness of $2\ \mu\text{m}$ and a width of $4\ \mu\text{m}$,  which is encompassed in a cold rectangular chamber with a dimension of $10\times8\ \mu\text{m}^2$ and a wall temperature of 500~K. The beam with a temperature of 300~K is placed $1\ \mu\text{m}$ away from the left and bottom walls of the enclosure. Gas is filled between the beam and chamber. Unlike the continuum flow where the flow velocity is zero and the temperature is governed by the Fourier's heat conduction law, at rarefied conditions, the temperature inhomogeneity induces anisotropic momentum transfer that in turn produces pressure gradient and bulk gas flow. Due to the asymmetric geometry, momentum fluxes impinging on the beam surface are unbalanced, giving rise to a net Knudsen force~\cite{Passian2003,Nabeth2011}, which can be exploited for microstructure actuation and gas sensing~\cite{Strongrich2015}. Previous researches have shown that the thermal edge flow occurring near the boundary with sharp curvatures plays a critical roles in the formation of Knudsen force~\cite{Sone1997,Zhu2010}.

The DG-RED with $k=4$ is applied to solve the rarefied gas flows using the ITR-LOC scheme~\eqref{ITR2}. The truncated molecular velocity is set as $[-6,6]^3$. $96$ non-uniform velocity points are used to discretize $v_1$ and $v_2$, while 24 uniform points are used for $v_3$. For evaluation of collision terms, $32\times32\times24$ equidistant frequencies are employed. Fig.~\ref{BeamG}(b) illustrates the schematic of the unstructured triangular mesh, where more triangles are placed near the micro-beam. We first consider flows at  $Kn=0.13$, 1.30 and 12.96. The Knudsen numbers are calculated using $T_0= 400$~K and $H=1\ \mu$m. The total iterative steps and the CPU time to obtain the steady-state solutions vary for flows. For the same spatial and velocity discretization, the smaller the Knudsen number, the more iterative steps thus more CPU time are required. To obtain the solution of $Kn=12.96$ on 881 triangles, 84 steps are needed to reach the convergence criterion of $\max\{R_T,R_{\rho},R_{|\bm u|}\}<10^{-5}$, which cost 4.3 hours on 12 processors (OpenMP for parallelism).  

\begin{figure}
	\begin{centering}
		\begin{subfigure}
		\centering
		\includegraphics[width=0.32\textwidth]{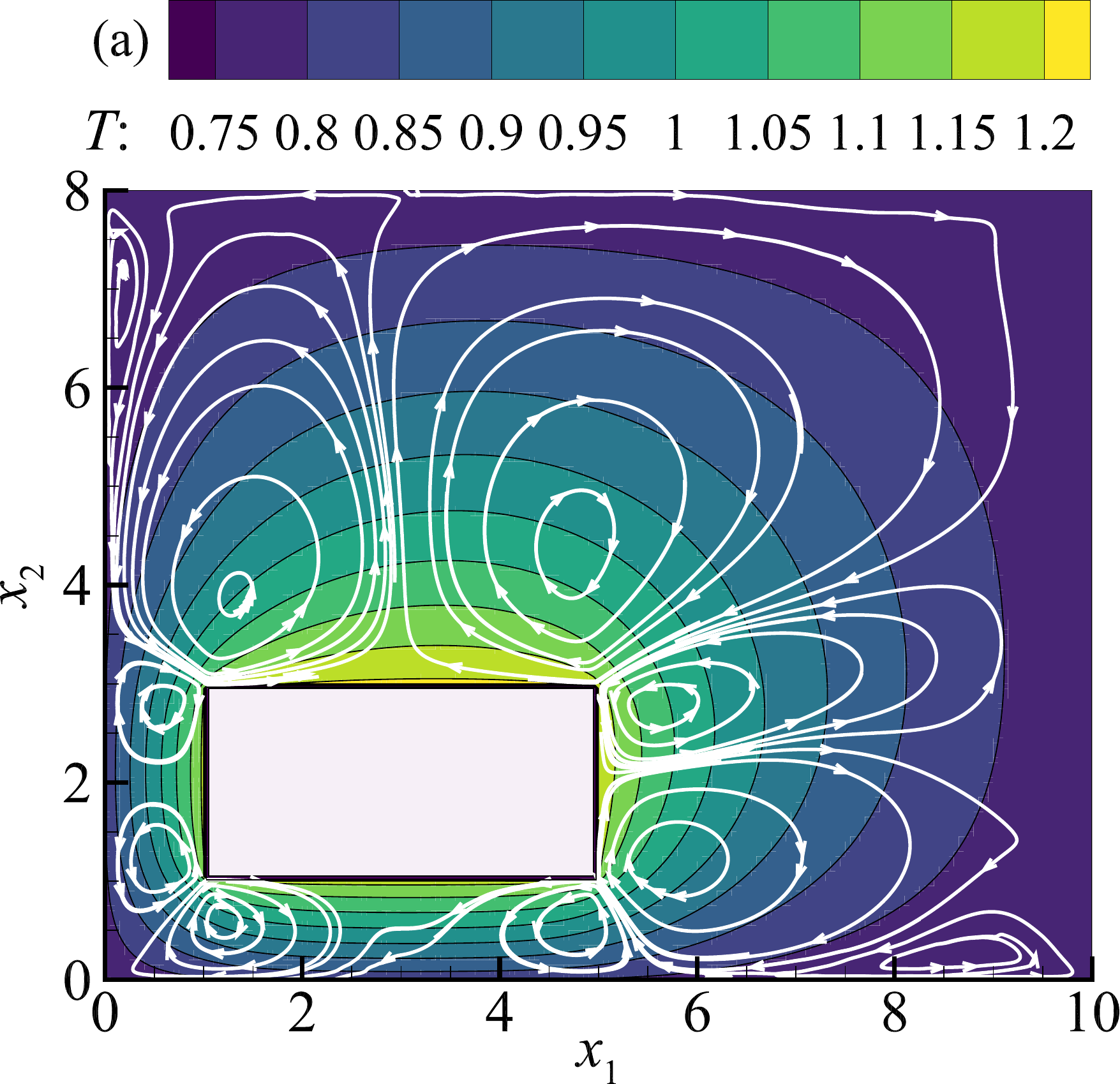}
		\end{subfigure}
		\begin{subfigure}
		\centering
		\includegraphics[width=0.32\textwidth]{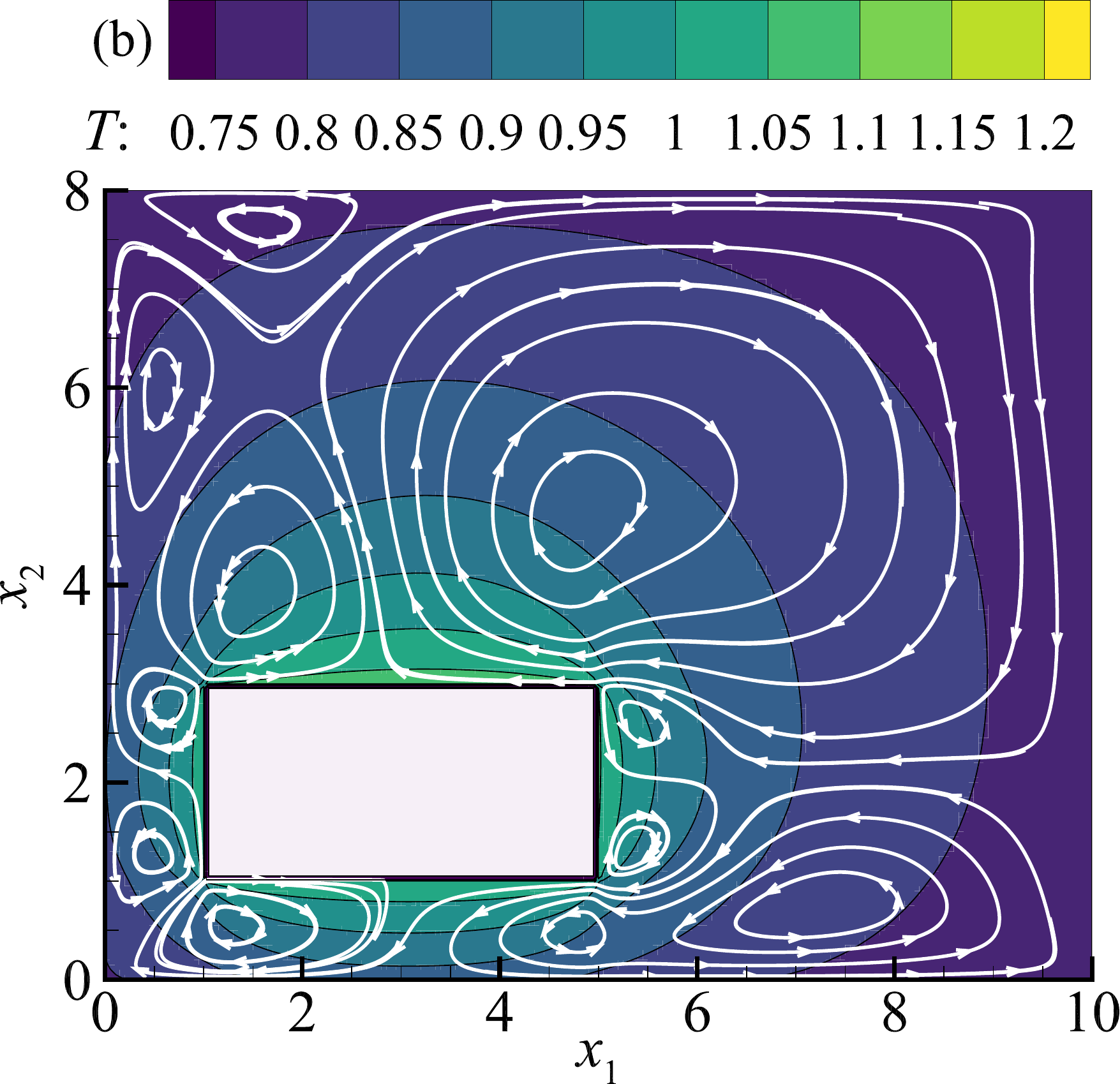}
		\end{subfigure}
		\begin{subfigure}
		\centering
		\includegraphics[width=0.32\textwidth]{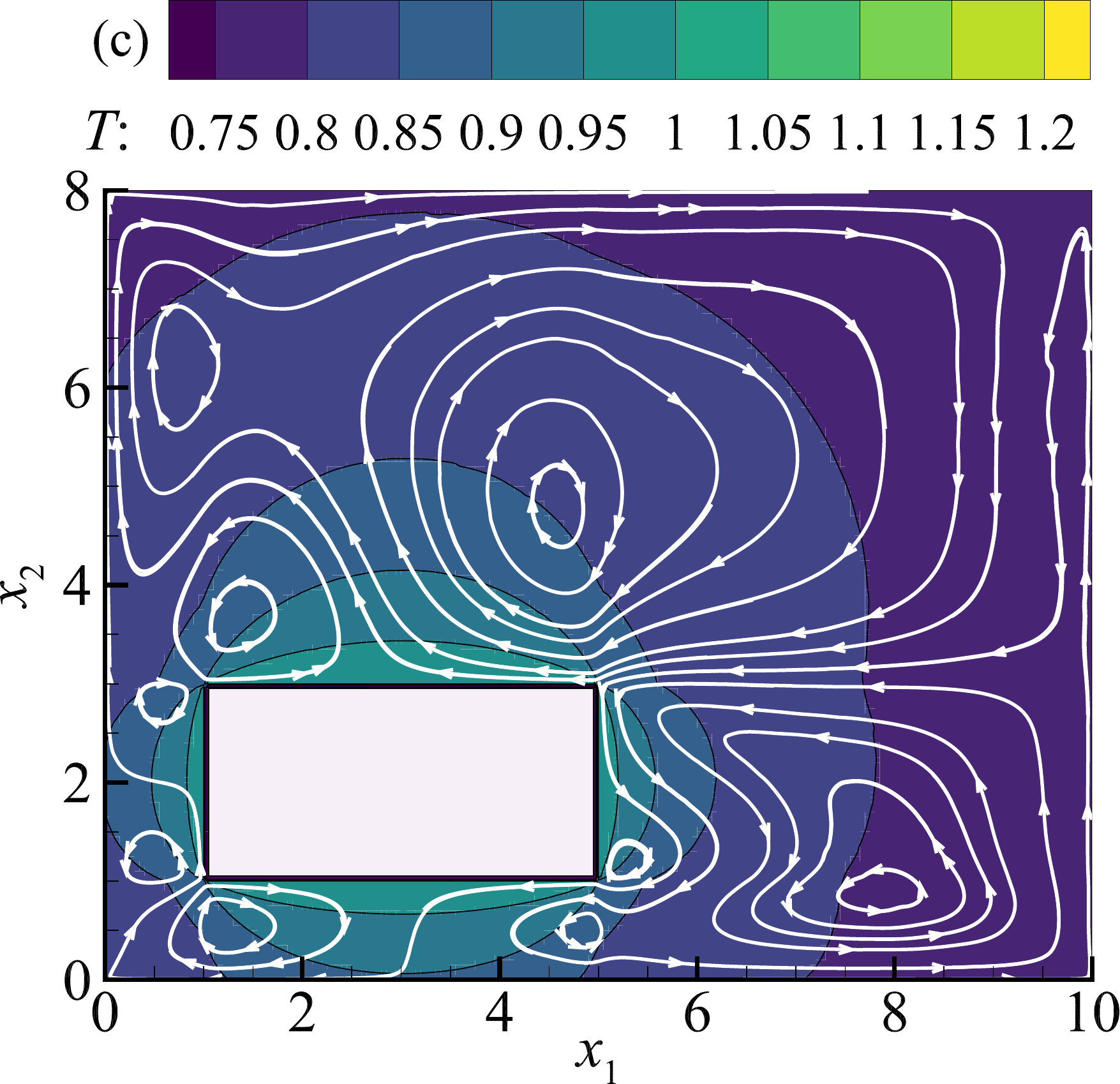}
		\end{subfigure}
		\par\end{centering}
	\caption{Temperature contours and streamlines in micro flow of argon gas with $\omega=0.81$. (a) $Kn=0.13$; (b) $Kn=1.30$; (c) $Kn=12.96$. The molecular velocity domain $[-6,6]^3$ is discretized by $96\times96\times24$ non-uniform grid points. 1290 triangles are used for flows of $Kn=0.13$ and $Kn=1.30$, while 881 triangles for case of $Kn=12.96$.}
	\label{BeamField}
\end{figure}

Figure~\ref{BeamField} shows the temperature contours and streamlines. It is observed that noticeable curls that originate at the corners of the beam emerge in the temperature contour lines at highly rarefied condition ($Kn=12.96$). However, in small Knudsen number flow, sufficient intermolecular collisions gradually smooth these curls when they propagate to the chamber. When the Knudsen number is small, at each surface of the beam, the thermal edge flows drive gas molecules from the corners to the surface centers and form a relatively high pressure region therein. Then, the high pressure results in the appearance of Poiseuille flows that promote gas flowing to the chamber. Due to the confinement of chamber walls, gas molecules finally return to the corners of the beam. Hence, eight localized vortices are observed in the flow field. When $Kn$ increases to 1.30, 3 more vortices are developed with one in the lower-right corner of the chamber and two in the upper-left corner of the chamber. As the degree of rarefaction further increases, the vortex in the lower-right corner of the chamber gradually dissolves the localized vortices near the right and bottom sides of the beam, which forms a large counter-clockwise vortex. Besides, the vortices in the region above the beam also start to melt together.

\begin{figure}
	\begin{centering}
		\includegraphics[width=0.85\textwidth]{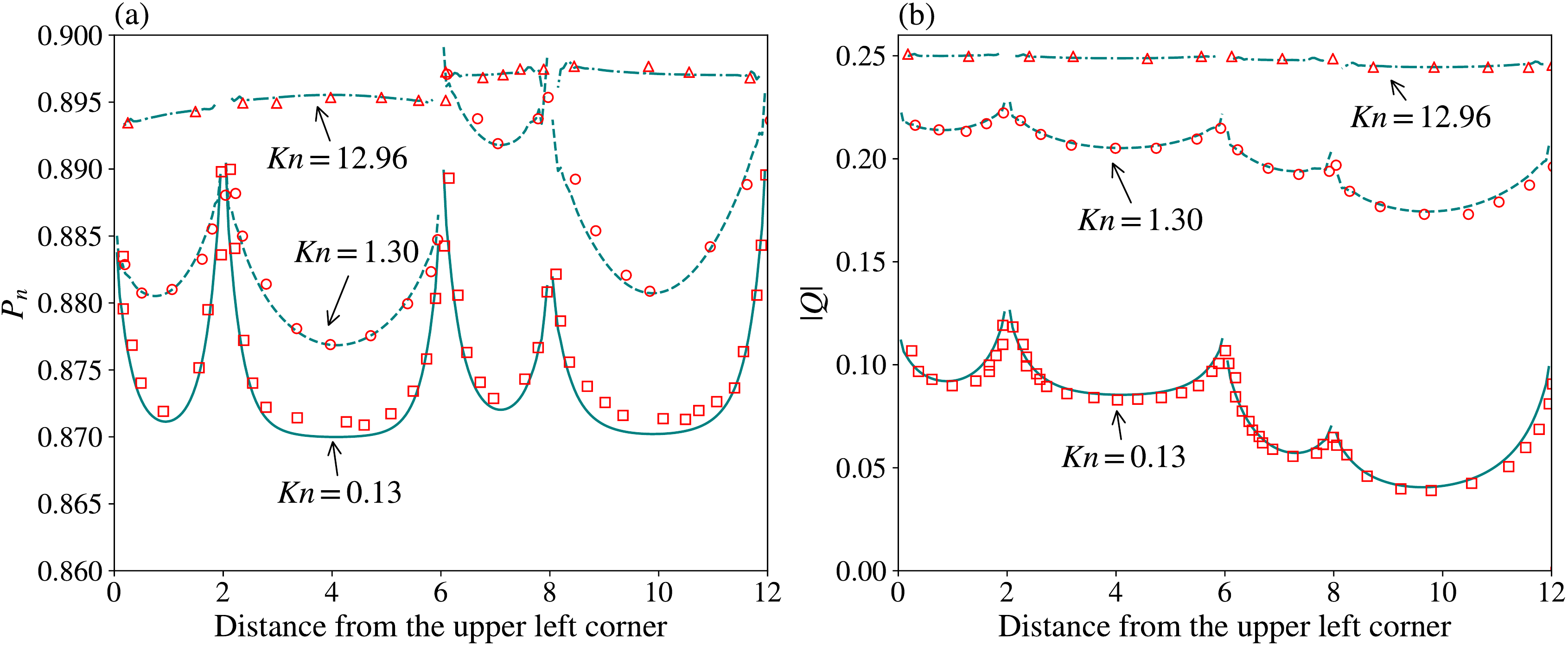}
		\par\end{centering}
	\caption{(a) normal stress and (b) magnitude of heat flux along the surface of hot beam. Lines are the DG-RED solutions and symbols are the DSMC result in Ref.~\cite{ZHU2017}. The horizontal axis represents the distance starting from the left-upper corner in a counter-clockwise direction. The gas is argon with $\omega=0.81$.}
	\label{BeamPQ}
\end{figure}

Figure~\ref{BeamPQ} illustrates the normal stress (pressure) $P_\text{n}$ and the magnitude of heat flux $|\bm Q|$ distributed on the surfaces of the hot beam, where $P_\text{n}$ is calculated as $P_\text{n}=\bm n_\text{w}\cdot\bm P\cdot\bm n_\text{w}$ with $\bm n_\text{w}$ denoting the outward unit normal vector of the beam surfaces. The DSMC solutions in Ref.~\cite{ZHU2017} are included for comparison, where good agreement can be observed. It can be seen that the more rarefied flow the larger $P_\text{n}$. This is due to the fact that momentum fluxes are enhanced when fewer intermolecular collisions are involved. Moreover, heat transfer is also strengthened by the non-equilibrium effect. The unbalance of $P_\text{n}$ on the surfaces mainly contributes to arising of the Knudsen force (the shear stress component is smaller than the normal one by two orders of magnitude). It is observed that $P_\text{n}$ on the top (right) surface of the beam is greater than that on the bottom (left) surface, thus both the horizontal and vertical components of the resultant force point to the negative directions of axes.  

\begin{figure}[t]
	\begin{centering}
		\includegraphics[width=0.95\textwidth]{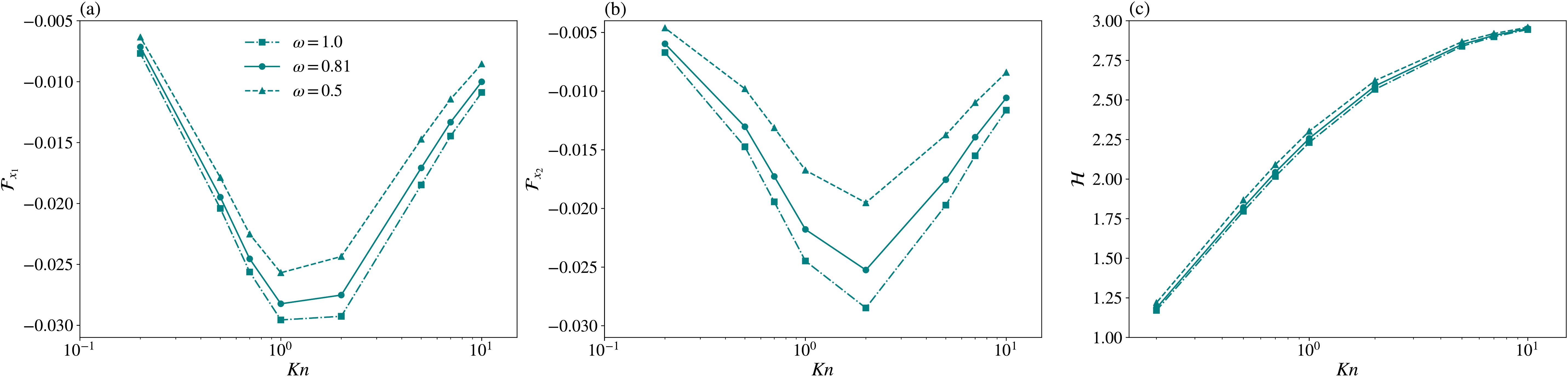}
		\par\end{centering}
	\caption{Resultant (a)-(b) force $[\mathcal{F}_{x_1},\mathcal{F}_{x_2}]^\mathrm{T}=-\oint_{\partial\Omega_\text{b}}\bm P\cdot\bm n_\text{w}\mathrm{d}\Upsilon$ acting on the hot beam; (c) total heat $\mathcal{H}=\oint_{\partial\Omega_\text{b}}\bm Q\mathrm{d}\Upsilon$ releasing by the hot beam. Solutions for argon of $\omega=0.81$, hard-sphere molecules of $\omega=0.5$ and Maxwell molecules of $\omega=1.0$ at $Kn$ ranging from 0.2 to 10 are compared. The scheme of DG-RED of $k=4$ combing with the ITR-LOC iteration~\eqref{ITR2} is applied. The molecular velocity domian $[-6,6]^3$ is discretized by $96\times96\times24$ non-uniform grid points. 1290 triangles are used for all Knudsen numbers.} 
	\label{BeamFH}
\end{figure}

The resultant force $\mathcal{F}$ acting on the hot beam and total heat $\mathcal{H}$ releasing from the hot beam at $Kn$ ranging from 0.2 to 10 are plotted in Fig.~\ref{BeamFH}, where results for argon molecules with $\omega=0.81$, hard-sphere molecules with $\omega=0.5$ and Maxwell molecules with $\omega=1.0$ are compared. The force and heat are calculated from integration as

\begin{equation}
[\mathcal{F}_{x_1},\mathcal{F}_{x_2}]^\mathrm{T}=-\oint_{\partial\Delta_\text{b}}\bm P\cdot\bm n_\text{w}\mathrm{d}\Upsilon,\quad\mathcal{H}=\oint_{\partial\Delta_\text{b}}\bm Q\mathrm{d}\Upsilon,
\end{equation} 
where, $\partial\Delta_\text{b}$ represents the surfaces of the beam. It is observed that the magnitude of Knudsen force first rises and then falls against the Knudsen number. The maximum magnitude of Knudsen force occurs around $Kn=2.0$. The total heat always increases with increasing $Kn$. The variation of Knudsen force can be ascribed to the development and competition of the localized thermal flows described above. When $Kn$ is small, i.e. the non-equilibrium effect is light, the variation of pressure on each beam surface is small and about the same magnitude, hence the Knudsen force is weak. As the Knudsen number increases, the strength of local flows are enhanced, and the more spacious spaces on the top and right of the beam allow formations of bigger vortices, which drive more gas molecules from the upper- and lower-right corners of the chamber to the center of the right surface of the beam, causing the pressure there to be larger than that near the left beam surface. On the other hand, the counter-clockwise vortex originating from the lower-right corner of the chamber penetrates into the bottom of the beam and efficiently takes gas molecules away from there. This causes the pressure near the bottom surface of the beam to be lower than that on its top surface. Therefore, the magnitudes of the horizontal and vertical components of Knudsen force both become larger. As the Knudsen number further increases, the thermal flows are further strengthened. The large vortex on the top surface of the beam starts to swallow the small vortices near the upper-left corner of the chamber, while the large vortex at the lower-right corner of the chamber begins to dissolve the small vortices on the right surface of the beam. The formations of two giant vortices release some pressure on the top and right surfaces of the beam, thus the magnitude of Knudsen force falls down. The profiles of $\mathcal{F}$ and $\mathcal{H}$ for $\omega=0.81$ always lie between the ones for $\omega=0.5$ and $\omega=1$.

\subsection{2D thermal cavity flow}

We also test the thermal cavity flow induced by temperature gradients at wall, which was recently used to verify an explicit DG Boltzmann solver by comparing with DSMC results~\cite{JAISWAL2018}. In this section, we intend to provide accurate results for this flow that may serve as benchmark solutions, when the Knudsen numbers are $Kn=0.1$, 0.5 and 1.

The computational domain is $1\times1$ square partitioned by structured triangular mesh as shown in Fig.~\ref{Cavity1}(a). The left and right walls are maintained at constant temperature $T_\text{C}$, while the bottom and top walls have varied temperature given by:
\begin{equation}
T\left(x_1,x_2=0\ \text{or}\ 1\right)=\begin{cases}
2\left(T_\text{H}-T_\text{C}\right)x_1+T_\text{C},\quad\quad\quad\quad\ \ x_1\leq0.5,\\
-2\left(T_\text{H}-T_\text{C}\right)x_1+2T_\text{H}-T_\text{C},\quad x_1>0.5,
\end{cases}
\end{equation}
where $T_\text{C}$ and $T_\text{H}$ are set as 263 K and 283 K, respectively.

The argon gas with viscosity index $\omega=0.81$ is initialized at the reference temperature of $T_0=273$ K. For all cases, the molecular velocity domain is chosen as $[-6,6]^3$, which is discretized by $72\times72$ non-uniform points in the $v_1$ and $v_2$ directions, and 24 uniform points in the $v_3$ direction. The corresponding frequency space, however, are discretized by $32\times32\times24$ equidistant frequencies for evaluation of the collision operator.

\begin{figure}
	\begin{centering}
	\begin{subfigure}
		\centering
		\includegraphics[width=0.33\textwidth]{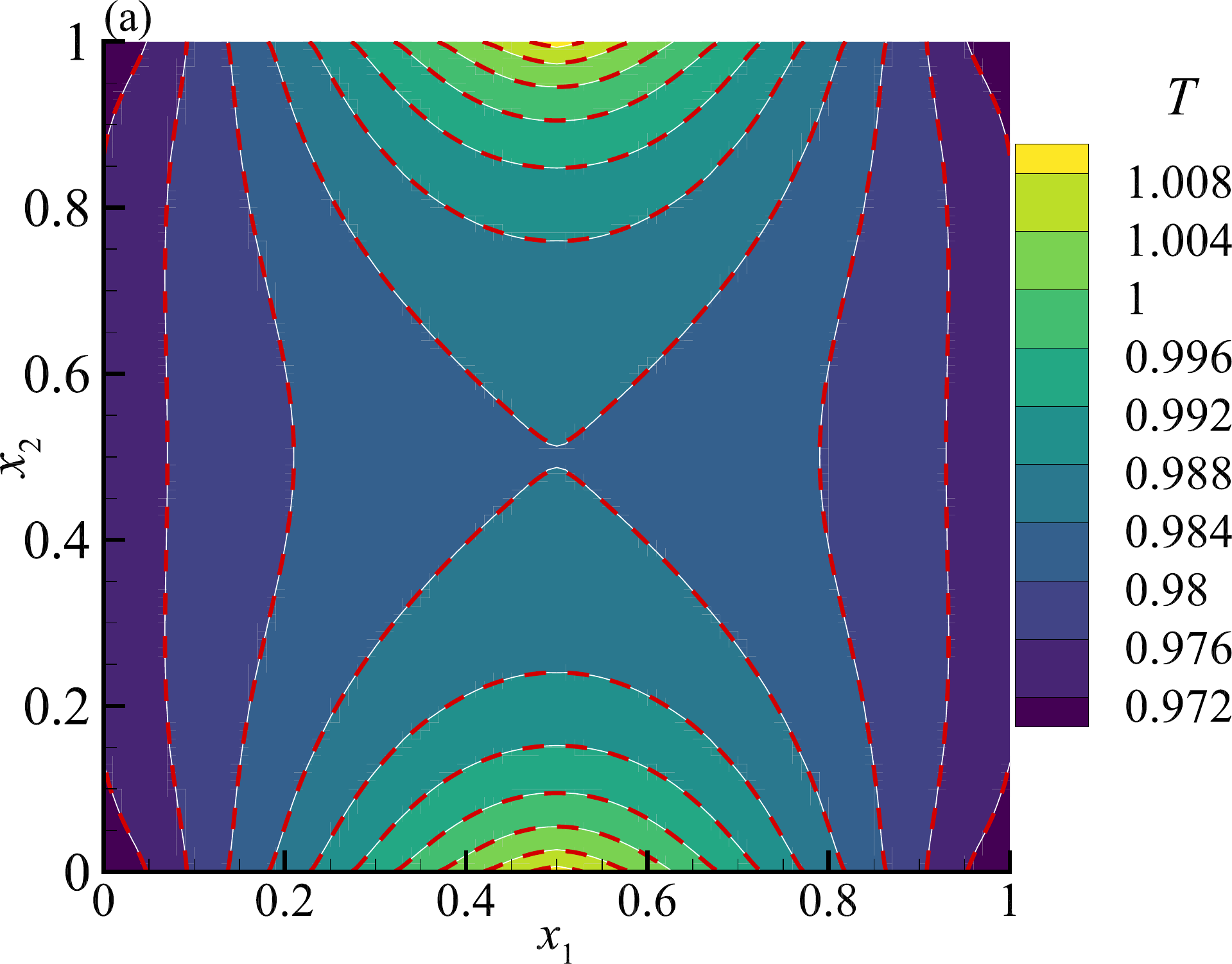}
	\end{subfigure}
	\begin{subfigure}
		\centering
		\includegraphics[width=0.335\textwidth]{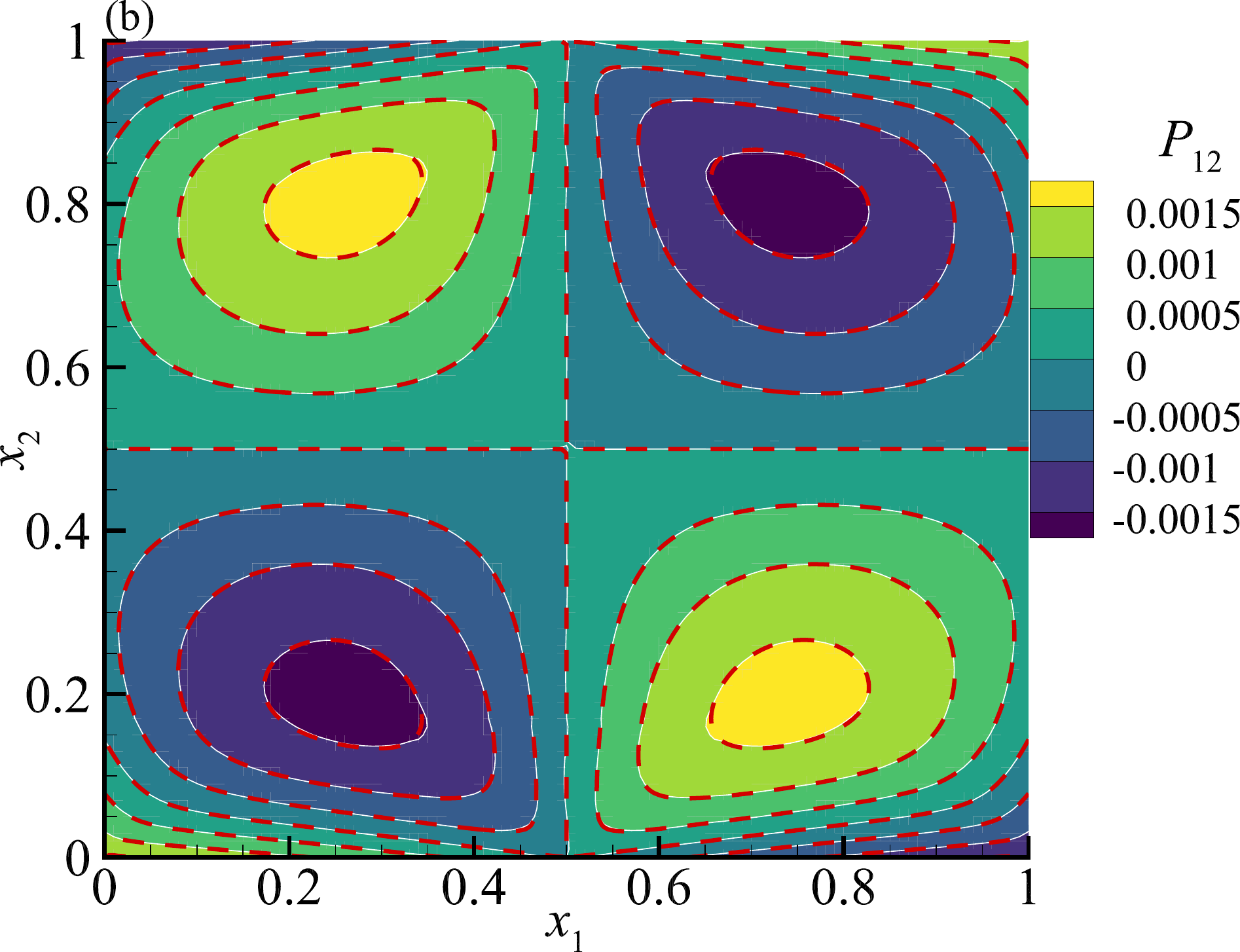}
	\end{subfigure}
	\begin{subfigure}
		\centering
		\includegraphics[width=0.27\textwidth]{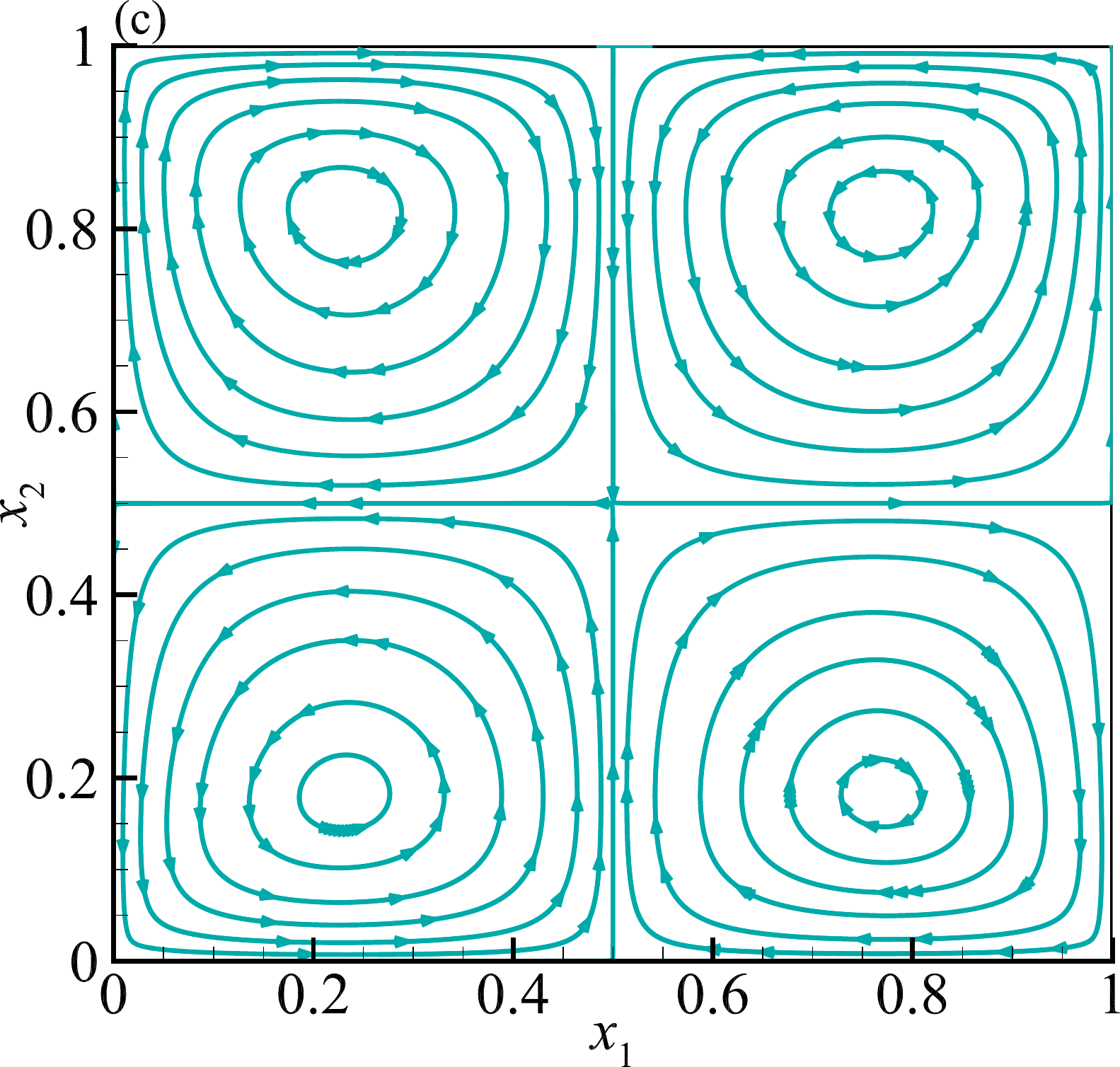}
	\end{subfigure}
		
	\par\end{centering}{}
	\caption{Comparison of the DG-RED and the FDM on the thermal cavity flow induced by the temperature gradients at wall when $Kn=0.5$. Contours of the temperature $T$ and shear stress $P_{12}$ are shown in (a) and (b), respectively, where the  solid lines with background are solutions of the DG-RED, while 
	the red dashed lines are the FDM results. (c) Stream lines. The DG-RED solutions are obtained with $k=4$ and 72 uniform triangles. The molecular velocity domain $[-6,6]^3$ is discretized by $72\times72\times24$ non-uniform grid points.}
	\label{ThermalCavity1}
\end{figure}

For verification of the DG results, the FDM results serve as reference solutions. In order to ensure accuracy of the FDM, $201\times201$ equidistant grid points are employed for the spatial discretization. Further refinement of both the velocity and spatial girds would only improve the solution by a magnitude no more than 0.5\%. The DG-RED scheme with $k=4$ is used to solve the flows on 72 triangles. The ITR-LOC iteration~\eqref{ITR2} is applied. Figure~\ref{ThermalCavity1} illustrates the dimensionless temperature and shear stress contours, as well as the streamlines for flow at $Kn=0.5$. The DG-RED steady-state solution presented here costs about 34 iterative steps and 0.96 hour CPU time on single processor. It is observed that high flow temperatures occur near the centers of bottom and top walls due to heating from the walls, while low temperatures appear in the four corners. The tangential temperature gradients near the walls lead to the thermal creep flows, where gas molecules along the bottom and top walls move from the colder regions towards the hotter ones. Due to the confinement of vertical walls, 4 vortexes are generated: two at the lower left and upper right quarters rotate counter-clockwise and the other two rotate clockwise. As a consequence, the maximum shear stresses appear at the centers of  clockwise vortices, while the minimum ones occur at the centers of counter-clockwise vortices. The flow patterns at $Kn=0.1$ and $Kn=1.0$ are similar. 

\begin{figure}[t]
	\begin{centering}
		\includegraphics[width=0.9\textwidth]{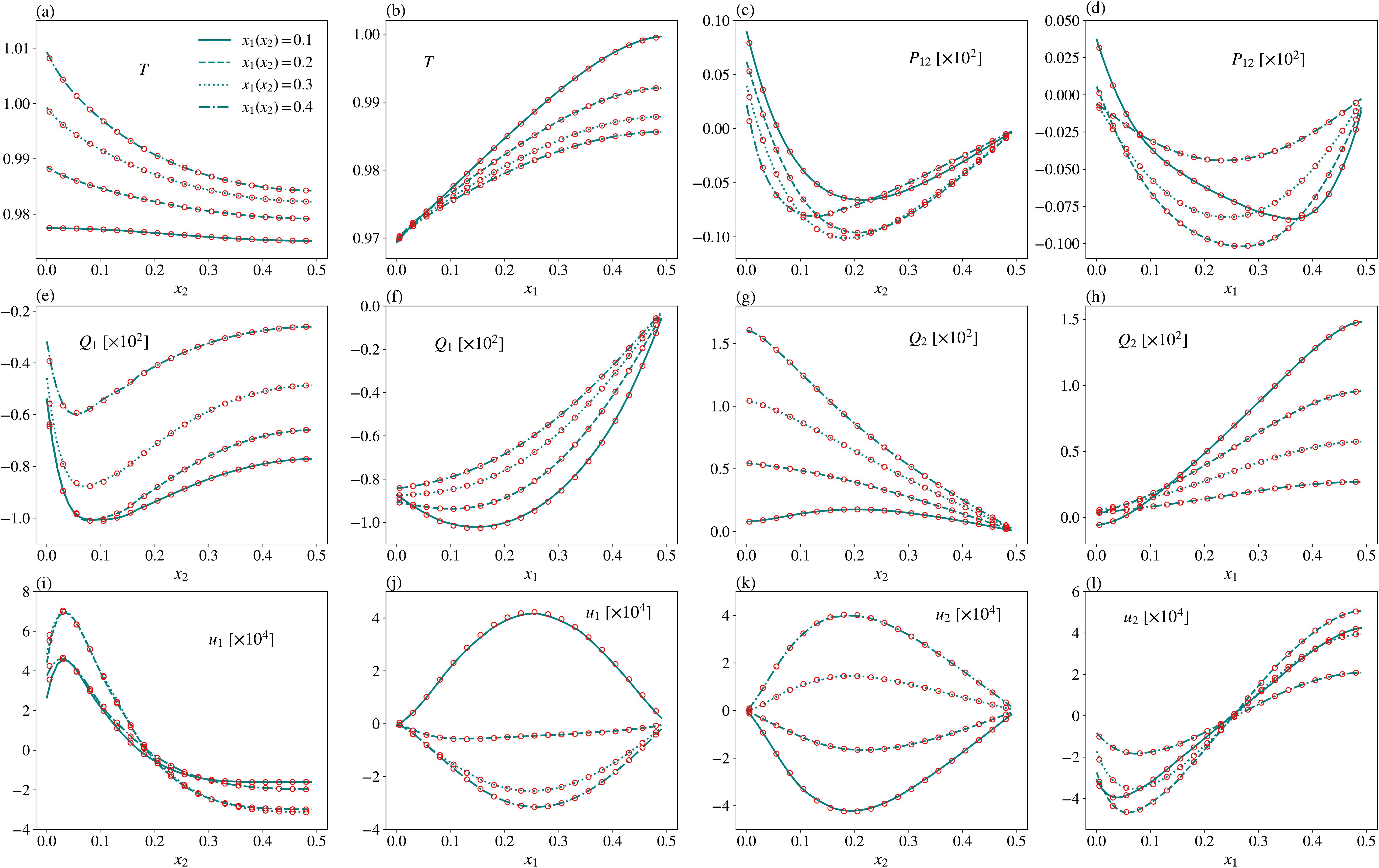}
		\par\end{centering}
\caption{Comparison of the DG-RED (solid lines) and the FDM (circles) on thermal cavity flow induced by temperature gradients at wall when $Kn=0.1$. The first and third columns are flow properties along vertical lines at $x_1=0.1$, 0.2, 0.3 and 0.4, while the second and forth columns are flow properties along horizontal lines at $x_2=0.1$, 0.2, 0.3 and 0.4. The DG-RED solutions are obtained with $k=4$ and 72 uniform triangles.} 
	\label{ThermalCavity21}
\end{figure}

\begin{figure}[t]
	\begin{centering}
		\includegraphics[width=0.9\textwidth]{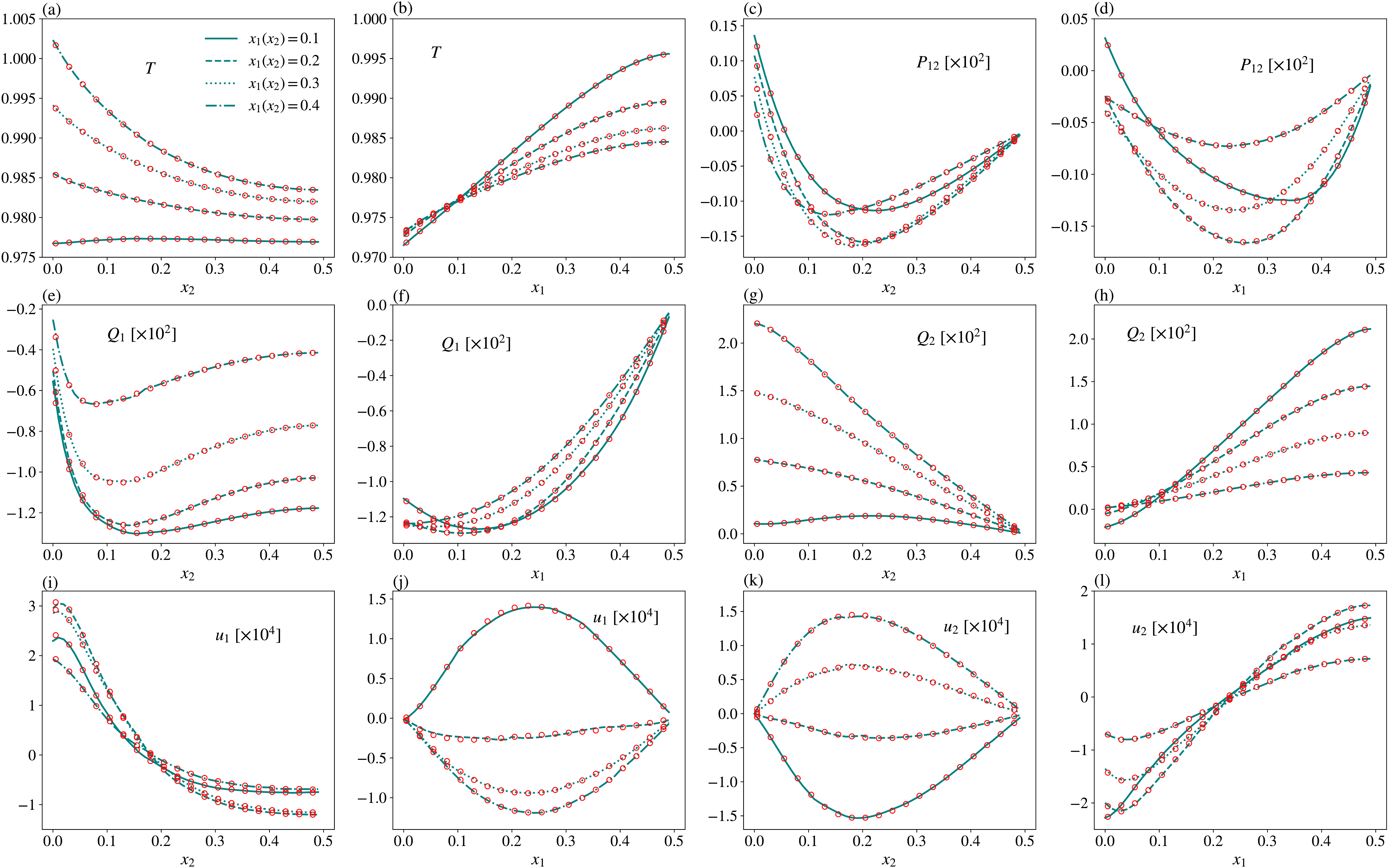}
		\par\end{centering}
	\caption{Comparison of the DG-RED (solid lines) and the FDM (circles) on thermal cavity flow induced by temperature gradients at wall when $Kn=0.5$.  The first and third columns are flow properties along vertical lines at $x_1=0.1$, 0.2, 0.3 and 0.4, while the second and forth columns are flow properties along horizontal lines at $x_2=0.1$, 0.2, 0.3 and 0.4. The DG-RED solutions are obtained with $k=4$ and 72 uniform triangles. } 
	\label{ThermalCavity22}
\end{figure}

\begin{figure}[t]
	\begin{centering}
		\includegraphics[width=0.9\textwidth]{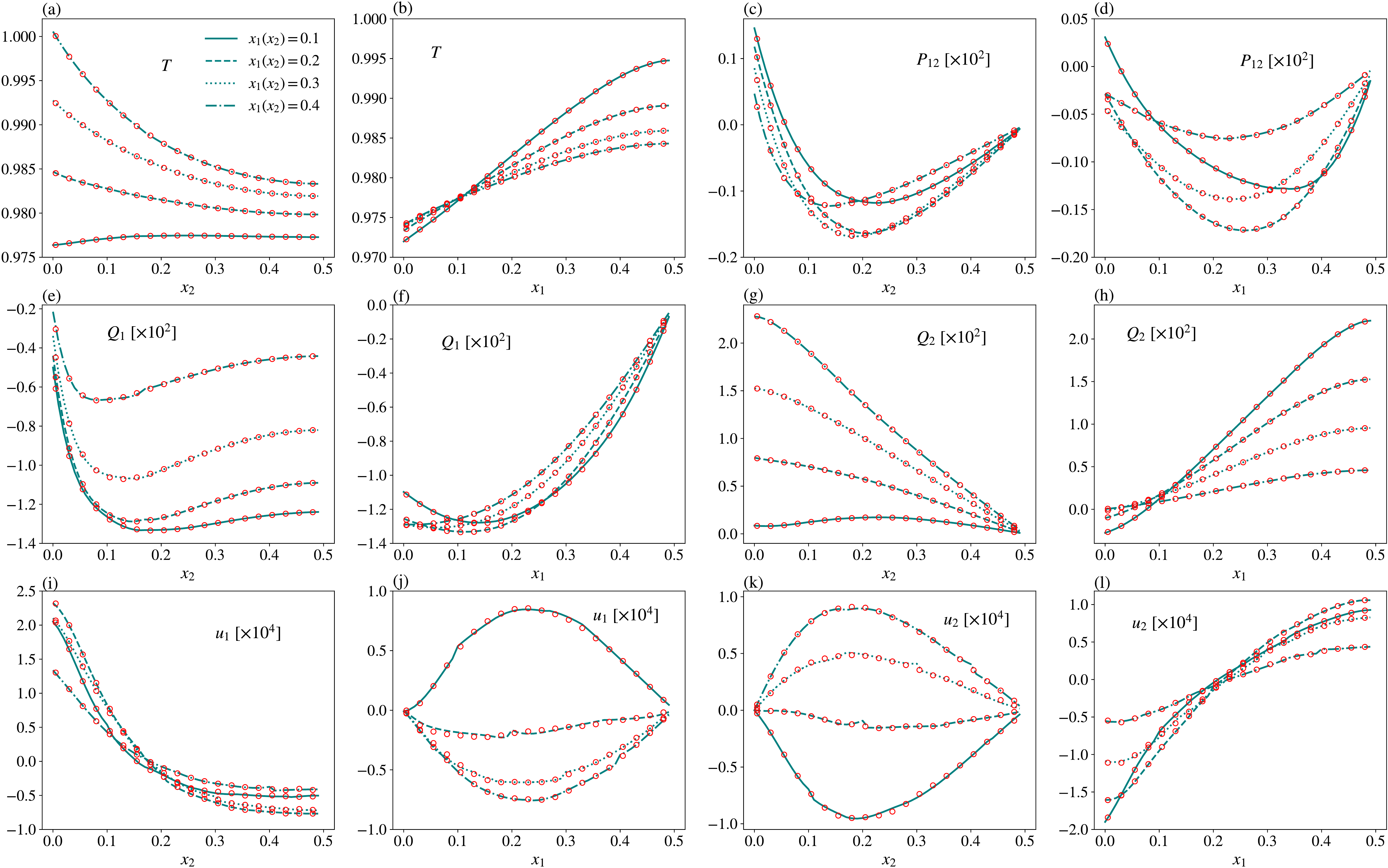}
		\par\end{centering}
	\caption{Comparison of the DG-RED (solid lines) and the FDM (circles) on thermal cavity flow induced by temperature gradients at wall when $Kn=1.0$. The first and third columns are flow properties along vertical lines at $x_1=0.1$, 0.2, 0.3 and 0.4, while the second and forth columns are flow properties along horizontal lines at $x_2=0.1$, 0.2, 0.3 and 0.4. The DG-RED solutions are obtained with $k=4$ and 72 uniform triangles.}
	\label{ThermalCavity23}
\end{figure}

Figure~\ref{ThermalCavity21} illustrates the variations of temperature $T$, shear stress $P_{12}$, horizontal (vertical) heat fluxes $Q_1$ ($Q_2$) and horizontal (vertical) flow velocities $u_1$ ($u_2$) along selected horizontal and vertical lines for rarefied gas flow when $Kn=0.1$; those for $Kn=0.5$ and $Kn=1$ are plotted in Figs.~\ref{ThermalCavity22} and~\ref{ThermalCavity23}, respectively. Due to the symmetry of flow field, results are only shown within the lower left quarter of the computational domain. It is found that from the regions near solid walls to the flow field center, the gas temperature increases along horizontal lines, while decreases along vertical lines. However, along both the horizontal and vertical directions, the shear stress first drops to the local minimum values then rises back to zero. The variations of horizontal heat flux are similar as those of shear stress, while the changes of the vertical component of heat flux are in accordance with the variations of gas temperature. The variations of bulk flow velocity are more complicated. Along the vertical lines, the horizontal velocity $u_1$ first increases to the local peaked values and then falls to the minimums. Along the horizontal lines near the bottom wall, $u_1$ is positive and has a local maximum at $x_1=0.25$, while in the regions away from the bottom wall, $u_1$ becomes negative and has a local minimum at $x_1=0.25$. Similarly, near the left lateral wall, the vertical velocity $u_2$ is negative and gradually changes its sign and reaches the local maximal values when approaching to the field center along the horizontal lines. For all flow properties, agreement between the DG-RED and the FDM results is pretty good. It is also interesting to note that, as the degree of rarefaction increases, the maximum values of temperature decrease since the intensity of gas-gas/gas-wall interactions becomes weaker. On the other hand, the maximum value in magnitudes of heat fluxes $|\bm Q|$ occurring near the centers of the bottom and top walls becomes larger, due to the larger temperature jump in high rarefied gas.

\section{Conclusions}\label{Concludsion}

In summary, we have developed a high-order discontinuous Galerkin discretization to solve the Boltzmann equation with full collision operator. The proposed numerical scheme is based on the classical discrete velocity method. At each discrete velocity grid points, the velocity distribution function is approximated in the piecewise polynomial spaces of degree up to 4 in the spatial space. Concerning the resolution of the Boltzmann collision operator, we rely on the Carleman-representation-based Fourier techniques, which can preserve mass and momentum and energy with spectral accuracy. Due to incorporation of the DG discretization and the fast spectral method, the computational complexity to evaluate the collision operator is of the order of $O\left(K^2M_\text{el}M^2_\text{qua}\bar{N}^3\log\bar{N}+2K^3M_\text{el}\bar{N}^3\right)$, which can be significantly large when high order approximating polynomial is used. Based on the nodal DG approximating, we have proposed a reduced DG discretization for the collision operator, which can reduce the computational complexity by $K$ times of magnitude.

Implicit iterative scheme scheme is employed to find the steady-state solution. At each iterative step, the DG discretization results in a system of linear equations for the degrees of freedom of velocity distribution functions on each spatial element. Since the first-order upwind principle is applied to approximate fluxes on the faces of spatial elements, the local linear equations only couple the unknowns on the immediate neighboring elements in the upwind side. Due to the fact that the direction of molecular velocity is fixed after discretization of the molecular velocity space, we have successfully employed the sweeping technique to sequentially solve the local linear systems, which avoids solving large sparse linear systems for all elements that are extremely expensive in both memory and CPU time when a large number of discrete velocities are required.

Five different test cases including hypersonic flows, as well as shear-driven and thermal-driven low-speed flows have been presented to show accuracy and capability of the proposed method. Several conclusions are summarized through the performance analysis:
\begin{itemize}

\item The implicit iterative scheme has no restriction on time step by CFL condition. The DG schemes with different order of approximating polynomials can obtain steady-state solution of the same order of accuracy within same number of iterative steps. Thus, the higher-order the discretization, the fewer spatial elements thus less CPU time is required. 

\item Compared to the full DG discretization in the collision operator, the proposed reduced DG approximation preserves the accuracy of the numerical scheme even for highly non-equilibrium flows, and significantly reduces the computational cost. To obtain the results with the same order of accuracy, the higher degree of approximation polynomial, the more the saving of CPU time in the reduced DG approximation.

\item Based on the same fast spectral method for the approximation of the Boltzmann collision operator, comparison with the finite difference method shows that the DG discretization is more efficient. Equipped with the implicit iterative scheme involving global mean collision frequency, the DG scheme can be faster than the finite difference method by one order of magnitude.

\item The implicit iterative scheme combining with the sweeping technique to sequentially solve the local linear systems on each spatial element preserves the stability of the DG scheme. Since in rarefied gas flow simulations the shock wave structure are resolved by fine spatial grids, the proposed method can solve hypersonic flows without any nonlinear limiter. 
\end{itemize}

The developed numerical method is straigtforward to be extended for the simulation of rarefied gas mixtures, where the velocity distribution function for each species is governed by its own Boltzmann equation. The Boltzmann equations for all constituents are coupled through pairwise collision operators. Thus, the computational complexity in resolving the collision terms via the FSM significantly increases as the number of gas species increases. In such situation, the advantage of using implicit DG method as well as the reduced calculation in collision operator will become more pronounced. Moreover, by incorporating more realistic intermolecular potentials such as the Lennard-Jones potential or even the ab initio potential based on quantum scattering~\cite{SHARIPOV2018797}, the developed scheme is ready to simulate a wide range of rarefied gas problems.

\section*{Acknowledgments}

This work is founded by the Engineering and Physical Sciences Research Council (EPSRC) in the UK under grant EP/R041938/1.

\appendix

\section*{Appendix}
\renewcommand{\theequation}{A.\arabic{equation}}

 Here, we present details of the DG formulation for the Boltzmann equation. The linear systems~\eqref{linear} to determine the solution of $f^{j'}$ on spatial element $\Delta_i$ are recalled here: 
\begin{equation}
\mathbf{A}^{i,j'}\mathbf{F}^{j'}_i+\mathbf{B}^{\text{ext},j'}=\mathbf{S}^{i,j'},,
\label{LPA}
\end{equation}
for $i=1,\dots,M_\text{el},\ j'=1,\dots,M$. 

We denote $F^{j'}_r=F_r(\bm v^{j'})$, $\Lambda^{j'}_p=\Lambda_p(\bm v^{j'})$ and $\Xi^{j'}_{p,r}=\Xi_{p,r}(\bm v^{j'})$ as values of the corresponding variables at each discrete velocity point, and $\mathbf{F}^{j'}_i=[F^{j'}_1,\dots,F^{j'}_r,\dots]^\mathrm{T}$ is the vector of degrees of freedom of $f^{j'}$ on $\Delta_i$. For ITR-LOC scheme, the coefficient matrices are:
\begin{equation}
\mathbf{A}^{i,j'}_{sr}=\frac{1}{2}\left(\bm v^{j'}\cdot\bm n+|\bm v^{j'}\cdot\bm n|\right)\langle\varphi_s,\varphi_r\rangle_{\partial\Delta_i}-\left(\bm v^{j'}\cdot\nabla\varphi_s,\varphi_r\right)_{\Delta_i}+\sum^{K}_{p=1}\left(\varphi_s,\varphi_p\varphi_r\right)_{\Delta_i}\Lambda^{j'}_p,
\end{equation}
\begin{equation}
\mathbf{B}^{\text{ext},j'}_{s}=\begin{cases}
\frac{1}{2}\left(\bm v^{j'}\cdot\bm n-|\bm v^{j'}\cdot\bm n|\right)\sum^{K}_{r=1}\langle\varphi_s,\varphi^{\text{ext}}_r\rangle_{\partial\Delta_i}F^{j'}_{r,\text{ext}},\quad\partial\Delta_i\not\subset\partial\Delta \\
\frac{1}{2}\left(\bm v^{j'}\cdot\bm n-|\bm v^{j'}\cdot\bm n|\right)\langle\varphi_s,b^{j'}\rangle_{\partial\Delta_i}, \quad\partial\Delta_i\subset\partial\Delta
\end{cases}
\end{equation}
\begin{equation}
\mathbf{S}_s=\sum^{K}_{p=1}\sum^{K}_{r=1}\left(\varphi_s,\varphi_p\varphi_r\right)_{\Delta_i}\Xi^{j'}_{p,r}
\end{equation}
where $\varphi^{\text{ext}}_r$ denotes the supporting polynomials on the neighboring element, from which $f_{\text{ext}}$ is obtained. For ITR-MEAN scheme, the coefficient matrices become:
\begin{equation}
\mathbf{A}^{i,j'}_{sr}=\frac{1}{2}\left(\bm v^{j'}\cdot\bm n+|\bm v^{j'}\cdot\bm n|\right)\langle\varphi_s,\varphi_r\rangle_{\partial\Delta_i}-\left(\bm v^{j'}\cdot\nabla\varphi_s,\varphi_r\right)_{\Delta_i}+\left(\varphi_s,\varphi_r\right)_{\Delta_i}\bar{\nu},
\end{equation}
\begin{equation}
\mathbf{B}^{\text{ext},j'}_{s}=\begin{cases}
\frac{1}{2}\left(\bm v^{j'}\cdot\bm n-|\bm v^{j'}\cdot\bm n|\right)\sum^{K}_{r=1}\langle\varphi_s,\varphi^{\text{ext}}_r\rangle_{\partial\Delta_i}F^{j'}_{r,\text{ext}},\quad\partial\Delta_i\not\subset\partial\Delta \\
\frac{1}{2}\left(\bm v^{j'}\cdot\bm n-|\bm v^{j'}\cdot\bm n|\right)\langle\varphi_s,b^{j'}\rangle_{\partial\Delta_i}, \quad\partial\Delta_i\subset\partial\Delta
\end{cases}
\end{equation}
\begin{equation}
\mathbf{S}_s=\sum^{K}_{p=1}\sum^{K}_{r=1}\left(\varphi_s,\varphi_p\varphi_r\right)_{\Delta_i}\left(\Xi^{j'}_{p,r}-\Lambda^{j'}_pF^{j'}_r\right)+\sum^{K}_{r=1}\left(\varphi_s,\varphi_r\right)_{\Delta_i}\bar{\nu}F^{j'}_r
\end{equation}

In this paper, nodal shape functions are chosen as the approximating polynomials. Integrals of the shape functions such as $\left(\varphi_s,\varphi_r\right)$, $\left(\nabla\varphi_s,\varphi_r\right)$, $\left(\varphi_s,\varphi_p\varphi_r\right)$ and $\langle\varphi_s,\varphi_r\rangle$ can be obtained analytically. To evaluate $\langle\varphi_s,b^{j'}\rangle$, the Gaussian rule is applied.



\bibliographystyle{elsarticle-num}
\bibliography{mybibfile}

\end{document}